\documentclass[12pt]{article}

\usepackage{epsf,epsfig,epstopdf}
\usepackage{graphics}

\usepackage[pdftex]{hyperref}

\usepackage{enumerate}

\usepackage{url}
\usepackage{amsmath,amssymb}
\usepackage{cite}

\usepackage{a4wide}
\numberwithin{equation}{section}

\def\braket#1{\mathinner{\langle{#1}\rangle}}

\begin{document}
\setlength{\unitlength}{1mm}

\allowdisplaybreaks
\thispagestyle{empty}

\begin{flushright}
{\small
ITP-UU-12/37\\
SPIN-12/34 \\
FR-PHENO-2012-33\\
14 November 2012}
\end{flushright}

\vspace{\baselineskip}

\begin{center}
\vspace{0.5\baselineskip}
\textbf{\Large\boldmath
 Finite-width effects on threshold corrections to squark and gluino production
}
\\
\vspace{3\baselineskip}
{\sc P.~Falgari$^a$, C.~Schwinn$^b$, C.~Wever$^a$}\\
\vspace{0.7cm}
{\sl ${}^a$Institute for Theoretical Physics and Spinoza Institute,\\
Utrecht University, 3508 TD Utrecht, The Netherlands\\
\vspace{0.3cm}
${}^b$ Albert-Ludwigs Universit\"at Freiburg, 
Physikalisches Institut, \\
D-79104 Freiburg, Germany }

\vspace*{1.2cm}
\textbf{Abstract}\\
\vspace{1\baselineskip}
\parbox{0.9\textwidth}{ 
We study the implication of finite squark and gluino decay widths for threshold resummation of squark and gluino production cross sections at the LHC.
We find that for a moderate decay width ($\Gamma/\bar{m} \lesssim 5\%$) higher-order soft and Coulomb corrections are appropriately described by NLL calculations in the zero-width limit including the contribution from bound-state resonances below threshold, with the remaining uncertainties due to finite-width effects 
of a similar order as the ambiguities of threshold-resummed higher-order calculations.
}
\end{center}

\newpage
\setcounter{page}{1}
\section{Introduction}

A recent development in the effort to improve theoretical predictions for
the pair production of squarks and gluinos at hadron colliders has
been the calculation of higher-order threshold corrections in QCD~\cite{Kulesza:2008jb,Kulesza:2009kq,Beenakker:2009ha,Beenakker:2010nq,Beenakker:2011sf,Langenfeld:2009eg,Langenfeld:2010vu,Langenfeld:2012ti,Beneke:2009rj,Beneke:2009nr,Beneke:2010da,Falgari:2012hx},
i.e. corrections that are enhanced in the limit of a small  relative velocity $\beta$ of the produced squark/gluino pair
\begin{equation}
\label{eq:thresh}
  \beta\equiv\sqrt{1-\frac{4\bar m^2}{\hat s}}\to 0,
\end{equation}
with the partonic centre-of-mass energy $\hat s$ and 
 the average  mass $\bar m$ of the sparticle pair.  
 These
threshold corrections are given by two contributions, Coulomb
corrections, that occur in powers of $\alpha_s/\beta$, and threshold
logarithms of the form $\alpha_s^n\ln^m\beta$.  
In the limit~\eqref{eq:thresh}, the expansion in the strong coupling constant must be reorganized, since both types of corrections can become of order one,
\begin{equation}
\label{eq:count-as}
\alpha_s\ln\beta\sim
1\;,\quad \frac{\alpha_s}{\beta}\sim 1.
\end{equation}
Resumming all enhanced contributions to all orders using eikonal
methods~\cite{Sterman:1986aj,Catani:1989ne,Kidonakis:1997gm,Bonciani:1998vc}
and non-relativistic field theory~\cite{Fadin:1987wz}, the cross section assumes the form
\begin{eqnarray}
\label{eq:nll-def}
\hat{\sigma}_{p p'} &\propto& \,\hat\sigma^{(0)}\, 
\sum_{k=0} \left(\frac{\alpha_s}{\beta}\right)^k \,
\exp\Big[\underbrace{\ln\beta\,g_0(\alpha_s\ln\beta)}_{\mbox{(LL)}}+ 
\underbrace{g_1(\alpha_s\ln\beta)}_{\mbox{(NLL)}}+
\underbrace{\alpha_s g_2(\alpha_s\ln\beta)}_{\mbox{(NNLL)}}+\ldots\Big]
\nonumber\\[0.2cm]
&& \,\times
\left\{1\,\mbox{(LL,NLL)}; \alpha_s,\beta \,\mbox{(NNLL)}; 
\alpha_s^2,\alpha_s\beta,\beta^2 \,\mbox{(NNNLL)};
\ldots\right\}.
\end{eqnarray}
For squark and gluino production, the corrections beyond
next-to-leading order~(NLO) can be large, becoming as large as the NLO
cross sections for the case of heavy gluinos.

In the calculations of threshold corrections the produced squarks and
gluinos are usually treated as stable, while in most realistic SUSY
models they are highly unstable and decay promptly.\footnote{We do not
  consider the case of stops and gluinos that are stable on collider
  timescales. For recent work on higher-order corrections in this case
see~\cite{Kauth:2009ud,Younkin:2009zn}.} In a more realistic treatment including
finite decay widths $\Gamma$, the threshold singularities are
smoothed~\cite{Fadin:1987wz} since the decay width sets the smallest
perturbative scale in the process.  This raises the question of whether the
resummation based on the counting~\eqref{eq:count-as} is appropriate
in the presence of the new scale $\Gamma$ and to which extent the
large higher-order threshold corrections found for stable squarks and
gluinos provide a reasonable approximation to the more realistic
unstable case.
One effect of the finite decay width is the emergence of a non-vanishing partonic cross section below the nominal production threshold $\hat s=4\bar m^2$, as  
discussed for stop-pair production in~\cite{Bigi:1991mi}, and more recently for gluino pair~\cite{Hagiwara:2009hq,Kauth:2011vg} and squark-gluino production~\cite{Kauth:2011bz}. In our previous combined resummation of threshold logarithms and Coulomb corrections for  squark and gluino pair-production~\cite{Beneke:2010da,Falgari:2012hx} the contribution from the region below the nominal threshold to the total production cross section has been taken into account in the zero-width limit, where the smooth invariant-mass distribution  turns into a series of would-be bound-state poles. 

As we will show in this note, for moderate decay widths
$\Gamma/\bar m\lesssim 5\%$, the  finite-width effects on
higher-order QCD corrections are small and within the error estimate of
the NLL cross sections with soft and Coulomb resummation and  bound-state corrections.
 This covers the relevant production channels in the MSSM, since larger
decay widths only arise for heavier gluinos with a large gluino-squark
mass splitting, in which  case
gluino production is suppressed compared to squark production.
Therefore the
treatment of NLL soft and Coulomb corrections in~\cite{Falgari:2012hx}
captures the dominant higher-order QCD effects also when the
instability of squarks and gluinos is taken into account. Furthermore,
the dominant effect of the finite decay widths arises at NLO, so the
impact on higher-order corrections is small.

It should be emphasized that the aim of our study is the assessment  of finite-width effects 
on QCD corrections beyond LO rather than a complete treatment  at leading and next-to-leading order. Studies 
of other aspects of finite squark and gluino decay widths include that
of non-resonant tree diagrams~\cite{Hagiwara:2005wg}; the accuracy of
the narrow-width approximation for small mass-splittings of the decaying
particle and its decay products~\cite{Berdine:2007uv,Uhlemann:2008pm}, the consistent
treatment of off-shell effects in Monte-Carlo programs~\cite{Gigg:2008yc}  
and NLO corrections to production and decay for the squark-squark production process~\cite{Hollik:2012rc}.

This article is organized as follows. In Section~\ref{sec:frame} we
specify the considered production and decay processes and review the
effective-theory treatment of unstable
particles~\cite{Beneke:2003xh,Beneke:2004km,Beneke:2007zg,Actis:2008rb}
and soft and Coulomb resummation near threshold~\cite{Beneke:2010da}.
In Section~\ref{sec:power-counting} we estimate the impact of the
screening of threshold corrections by the finite decay width as well as
the magnitude of non-resonant corrections.  In
Section~\ref{sec:application} we study the invariant mass spectrum for
squark and gluino production processes at the LHC and the total
production cross sections. We find that for moderate decay widths the
finite-width corrections to the total cross section are within the residual uncertainty
of the NLL calculation of~\cite{Falgari:2012hx}.

\section{Theoretical framework}
\label{sec:frame}
\subsection{Production and decay of  squarks and gluinos in SQCD}
In this note we consider the pair-production processes of squarks
$\tilde q$ and gluinos $\tilde g$ at hadron colliders in
supersymmetric QCD~(SQCD) that proceed through the partonic production
channels
\begin{equation}
  \label{eq:part-process}
  pp'\to \tilde{s}_1 \tilde{s}_2X\;,
\end{equation}
where $\tilde{s}_i\in\{\tilde q,\bar{\tilde q},\tilde g\}$ denote the two produced sparticles and $p,p'\in\{q,\bar q,g\}$ the initial-state partons.
The average  mass of the sparticle pair is given by
\begin{equation}
  \label{eq:mav}
\bar m=\frac{m_{\tilde{s}_1}+m_{\tilde{s}_2}}{2}.
\end{equation}
In general, the produced squarks and gluinos can decay via long
cascade decay chains to a final state of standard model particles and
the lightest supersymmetric particle (assuming the latter is stable on
detector time scales).  For this initial study of finite-width effects
we limit ourselves to squark and gluino decay in pure SQCD, i.e. we
neglect the decay width of the lightest coloured sparticle  and consider the
decay modes
\begin{align} \label{eq:qgdecay}
\tilde{q} &\rightarrow q\tilde{g}, \ \ \bar{\tilde{q}} \rightarrow \bar{q}\tilde{g}, \ \ m_{\tilde{q}}>m_{\tilde{g}}, \nonumber\\
\tilde{g} &\rightarrow q\bar{\tilde{q}}, \ \ \tilde{g} \rightarrow \bar{q}\tilde{q}, \ \ m_{\tilde{q}}<m_{\tilde{g}}.
\end{align}
We treat  
the masses of the different squark flavours as degenerate and equal to $m_{\tilde{q}}$, while the gluino mass equals $m_{\tilde{g}}\neq m_{\tilde{q}}$. 
The full one-loop SUSY-QCD 
partial widths of the above decay processes have been calculated in \cite{Beenakker:1996dw}. The NLO corrections to the LO partial widths were found to be of the order of $30-50\%$ and 
will not be given here. In order to estimate the finite-width effects we will employ  the LO tree level SUSY-QCD widths,
\begin{eqnarray} \label{eq:qgdecaywidth}
\Gamma(\tilde{q}\rightarrow q\tilde{g}) & = & \frac{\alpha_sC_Fm_{\tilde{q}}}{2}\Big(1-\Big(\frac{m_{\tilde{g}}}{m_{\tilde{q}}}\Big)^2\Big)^2, \ m_{\tilde{q}}>m_{\tilde{g}}, 
\nonumber\\
\Gamma(\tilde{g}\rightarrow q\bar{\tilde{q}},\bar{q}\tilde{q}) & = & \frac{\alpha_sn_fm_{\tilde{g}}}{2}\Big(1-\Big(\frac{m_{\tilde{g}}}{m_{\tilde{q}}}\Big)^{-2}\Big)^2, m_{\tilde{q}}<
m_{\tilde{g}}.
\end{eqnarray}
Since the numerical values are only used for illustration of
the finite-width effects, we only consider gluino decay into
light-flavour and bottom squarks that are treated as mass-degenerate, i.e. we set $n_f=5$.  
Therefore we neglect the decay into stops and the related additional
dependence on the parameters of the stop sector unless stated otherwise. 
The effect of gluino decay to stops on our results is discussed briefly in Section~\ref{sec:results}.
In the full MSSM, the lighter coloured sparticles are, of course,
unstable with respect to electroweak two- or three-body decays $\tilde q\to \chi q$ and
$\tilde g\to q\bar q \chi$ for charginos or neutralinos $\chi$,
which have been calculated in leading order
in~\cite{Baer:1986au,Baer:1990sc} and for which state-of-the-art
predictions are implemented
in~\cite{Porod:2003um,Porod:2011nf,Muhlleitner:2003vg,Djouadi:2006bz}.
These effects would be straightforward to include in our framework but
would introduce a dependence on the electroweak parameters of the
MSSM. Therefore we will use the tree-level SQCD decay widths~\eqref{eq:qgdecaywidth} for
the purpose of estimating the size of finite-width effects
in higher-order QCD corrections.
Since the electroweak decay widths are typically of the order
$\Gamma/\bar m< 1\%$, this should not affect our qualitative conclusions.
  We therefore consider
production and decay processes of squarks and gluinos at hadron
colliders of the type
\begin{equation}
\label{eq:example-process}
  pp'\to \tilde s_1\tilde s_2\to (\tilde s_1' p_1) \, (\tilde s_2' p_2),  
\end{equation}
with massless partons $p_{1/2}$.
Such four-body final states arise in SQCD for gluino pair production for the case $m_{\tilde g}>m_{\tilde q}$ or squark-(anti-)squark production for $m_{\tilde q}>m_{\tilde g}$.
The treatment of a three-body final state that arises for squark-gluino production in our approximation should be obvious. 
 In detail, the partonic 
 production and decay processes of squarks and gluinos for the two mass hierarchies are as follows:
\begin{align}
  m_{\tilde{g}}<m_{\tilde{q}}&:&
gg,q_i\bar{q}_i &\rightarrow \tilde{g}\tilde{g},\nonumber\\
&&q_ig &\rightarrow \tilde{q}\tilde{g} \rightarrow q\tilde{g}\tilde{g} ,
\nonumber\\
&&gg,q_i\bar{q}_j &\rightarrow \tilde{q}\bar{\tilde{q}}\rightarrow q\bar
q\tilde{g}\tilde{g},  \nonumber\\
&&q_iq_j& \rightarrow \tilde{q}\tilde{q} \rightarrow qq\tilde{g}\tilde{g},
 \nonumber \\
m_{\tilde{g}}>m_{\tilde{q}}&:&
gg,q_i\bar{q}_j &\rightarrow \tilde{q}\bar{\tilde{q}},\nonumber\\
&&q_iq_j &\rightarrow \tilde{q}\tilde{q},\nonumber\\
&&q_ig &\rightarrow \tilde{q}\tilde{g} 
\rightarrow \bar{q}\tilde{q}\tilde{q},\nonumber\\
&&gg,q_i\bar{q}_i &\rightarrow \tilde{g}\tilde{g} \rightarrow 
\bar{q}\bar{q}\tilde{q}\tilde{q}
\label{eq:decayprocesses}
\end{align}
where $i,\, j=u, \, d, \, s, \, c, \, b$ and all charge conjugated processes are understood. 

\begin{figure}[t!]
\centering
\includegraphics[width=0.7 \linewidth]{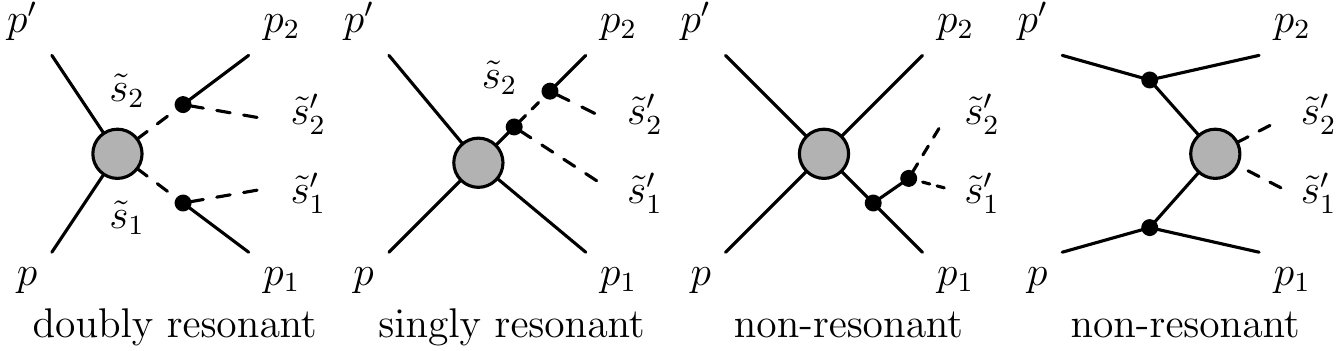}
\caption{Examples of Feynman-diagram topologies contributing to the generic production and decay
  process~\eqref{eq:example-process}. The second non-resonant diagram
  contains collinear singularities.} \label{fig:example-diagrams}
\end{figure}

The complete gauge-invariant set of Feynman diagrams for the
production of the three--or four-body final states of ``stable''
particles~\eqref{eq:decayprocesses} contain doubly-resonant diagrams,
i.e. diagrams where the particles indicated in the first step
of~\eqref{eq:decayprocesses} appear as internal lines, and singly- or
non-resonant diagrams, where only one or none of the unstable squarks
or gluinos is present. Examples of these topologies are given in
Figure~\ref{fig:example-diagrams}.  Beyond leading order, the
production processes cannot be strictly separated since, e.g. the
process $q_ig \rightarrow \bar{q}\tilde{q}\tilde{q}$ enters the real
NLO corrections to $\tilde{q}\tilde{q}$-production.  Furthermore, the
non-resonant diagrams can contain collinear singularities (e.g. the
last topology given in Figure~\ref{fig:example-diagrams}) that would
cancel against higher-order virtual corrections and require taking
kinematic cuts on the final state particles into account.

A full NLO QCD calculation of a process comparable to~\eqref{eq:example-process} including finite-width effects has been performed in the standard model for the case of $b\bar b W^+W^-$ production~\cite{Denner:2010jp,Bevilacqua:2010qb}, but not yet for an MSSM process. Very recently, a tool for  automatic calculation of MSSM processes at NLO has been presented in~\cite{GoncalvesNetto:2012yt}, but so far applied to the on-shell processes~\eqref{eq:part-process}.
Short of a full NLO calculation, a well-defined approximation method  for calculating radiative corrections to  production and decay processes is based on the pole expansion~\cite{Stuart:1991xk,Aeppli:1993rs}.
In this framework the total partonic  production cross sections for the processes~\eqref{eq:decayprocesses} is  written in the form
\begin{equation}
\label{eq:sigma-pole}
  \hat\sigma_{pp'}(\hat s)= 
 \hat\sigma_{pp'}^{\text{res}}(\hat s)
 + \hat\sigma_{pp'}^{\text{non-res}}(\hat s).
\end{equation}
The doubly-resonant cross-section $\hat\sigma^{\text{res}}$ is defined
in a consistent way by an expansion of the $S$-matrix around the
complex poles of the Dyson-resummed propagators of the unstable
particles~\cite{Stuart:1991xk,Aeppli:1993rs}.  At NLO it contains both
factorizable corrections to production and decay, as well as
non-factorizable corrections connecting production, propagation and
decay stages. Non-factorizable corrections related to the final state
cancel for the total cross
section~\cite{Fadin:1993dz,Melnikov:1993np}, but not for differential
distributions.  The fully differential factorizable corrections to
production and decay have recently been computed for squark-squark
production~\cite{Hollik:2012rc}.  For the remaining processes, at
present only the corrections to the total production cross
sections~\cite{Beenakker:1996ch} and decay
rates~\cite{Beenakker:1996dw} are available.  The ``non-resonant''
cross-section $\hat\sigma^{\text{non-res}}$ includes singly- and
non-resonant Feynman-diagrams as well as contributions from
doubly-resonant diagrams where one or both of the $\tilde s_1$ or
$\tilde s_2$ lines are far off-shell.

\subsection{Effective theory framework for unstable particles}
\label{sec:eft}

For the investigation of the interplay of higher-order threshold
corrections in QCD and finite squark and gluino lifetimes, we are
interested in the partonic cross sections for the
processes~\eqref{eq:decayprocesses} near the partonic production
threshold~\eqref{eq:thresh}.  For the precise definition of the
resonant and non-resonant contributions to the cross section we adopt
the effective-theory approach to unstable
particles~\cite{Beneke:2003xh,Beneke:2004km}, generalizing the
treatment of $W$-boson pair production at an electron positron
collider~\cite{Beneke:2007zg,Actis:2008rb}.  The aim of the effective-theory approach is to provide a precise prediction of the partonic
cross section for partonic centre-of-mass energies in the vicinity of
the threshold, $\hat s -4\bar m^2 \sim \bar m\Gamma$. This is achieved
by a simultaneous expansion in the quantity
\begin{equation}
  \delta=\frac{\hat s -4\bar m^2}{\bar m^2}\approx \beta^2 
\end{equation}
and the coupling constants. For power-counting purposes, we therefore set
\begin{equation}
\label{eq:def-delta}
  \beta\sim (\Gamma/\bar m)^{1/2}\sim\delta^{1/2}.
\end{equation}
The leading-order resonant production cross-section for $S$-wave production processes is of the form
\begin{equation}
\label{eq:s-wave}
   \hat\sigma^{\text{res}}_{pp'}(\hat s,\mu) \sim \alpha_s^2 \, \beta .
\end{equation}

Near the production threshold, the
processes~\eqref{eq:example-process} can be treated in an effective
theory where the light-partons are described by collinear fields
$\phi,\phi'$ in soft-collinear effective
theory~(SCET)~\cite{Bauer:2000yr,Bauer:2001yt,Beneke:2002ph} and the
pair-produced squarks and gluinos are described by non-relativistic
fields $\psi,\psi'$ in potential non-relativistic
QCD~(pNRQCD)~\cite{Brambilla:1999xf}. In the leading effective
Lagrangian, the collinear and non-relativistic fields interact only
through the exchange of soft gluons, which give rise to the
non-factorizable corrections mentioned above. This effective theory
has also been used for soft-gluon and Coulomb resummation
in~\cite{Beneke:2010da}, where more details  can be found.

\begin{figure}[t!]
  \centering
\includegraphics[width=0.5 \linewidth]{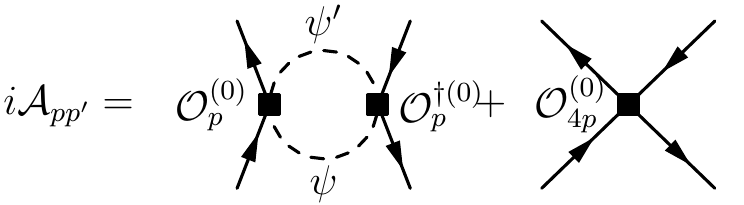}
  \caption{Diagrammatic representation of the leading resonant and non-resonant contribution to the forward-scattering amplitude in unstable particle effective theory.}
  \label{fig:eft-amp}
\end{figure}
\begin{figure}[t!]
  \centering
\includegraphics[width=0.7 \linewidth]{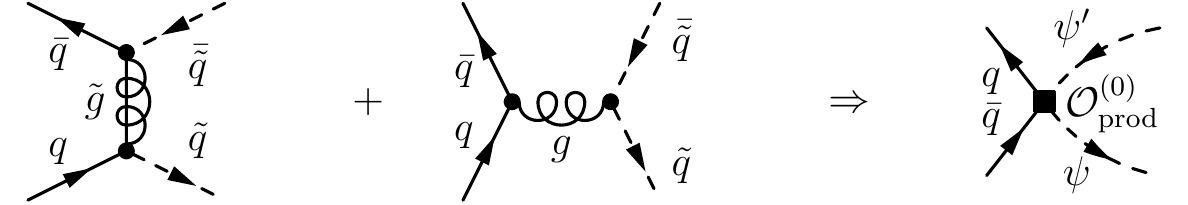}
\caption{
Computation of the matching coefficient of the production operator for squark-antisquark production from the quark-antiquark initial state.}
 \label{fig:match-res}
\end{figure}
\begin{figure}[t!]
  \centering
\includegraphics[width=0.7 \linewidth]{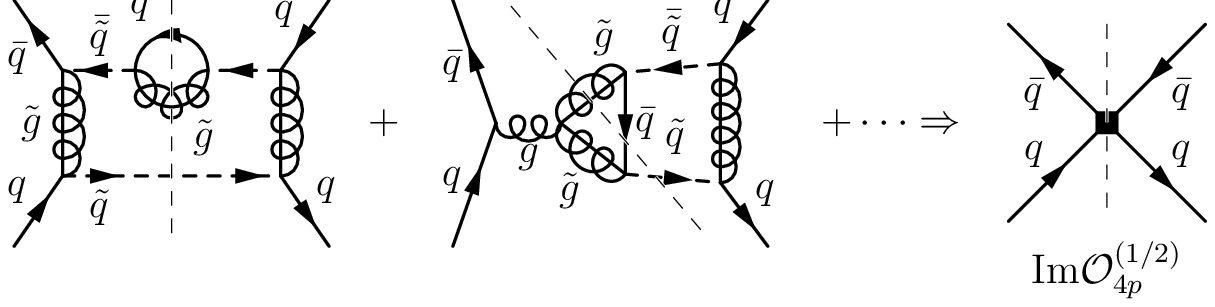}
  \caption{ Computation of the matching coefficient of the four-parton operator for squark-antisquark production and decay. The first diagram is an example of a double-resonant diagram, the second of the interference of a double-resonant and a single-resonant diagram.}
  \label{fig:match-nonres}
\end{figure}

In the EFT framework, the
total partonic cross section for the
process~\eqref{eq:example-process} is computed from the imaginary part
of the partonic forward-scattering amplitudes $\mathcal{A}_{pp'}$ of
the processes $pp'\to pp'$ 
which reads\footnote{As written, the forward-scattering amplitude is not
  infrared safe but requires mass factorization where matrix elements
  of the collinear fields are identified with parton distribution
  functions~(PDFs).}
\begin{equation}
\label{eq:eft-amp}
i \mathcal{A}_{pp'}(\hat s)|_{\hat s\sim 4\bar m^2}=  \int d^4 x \,
\braket{pp' |
T\left[i {\cal O}^\dagger_{\text{prod}}(0)
i{\cal O}_{\text{prod}}(x)\right]|pp'}
+ \braket{pp' |i {\cal O}_{4p}(0)|pp'} \, .
\end{equation}
The ``production operator'' ${\cal O}_{\text{prod}}= C_{\text{prod}}\,
\phi\phi'\,\psi^\dagger{\psi'}^\dagger$ describes the resonant
production of the $\tilde s_1\tilde s_2$-pair from $pp'$ collisions
while the expectation value of the four-parton operators $ {\cal
  O}_{4p}=C_{4p}\, \phi^\dagger{\phi'}^\dagger\phi\phi'$ describes the
non-resonant contributions.  The two terms in~\eqref{eq:eft-amp}
therefore correspond to the resonant and non-resonant cross sections
defined earlier in~\eqref{eq:sigma-pole}, see
Figure~\ref{fig:eft-amp}.  The cross section for a specific final
state $\tilde s_1' \tilde s_2' p_1p_2$ can be obtained by computing the
imaginary part of the forward-scattering amplitude using  unitarity
cuts and selecting only cuts corresponding to the desired final
state~\cite{Beneke:2007zg}.  At leading order, this simply amounts to
multiplying by the branching ratios $\text{BR}(\tilde q_1\to \tilde q_1'
p_1)$ and $\text{BR}(\tilde q_2\to \tilde q_2' p_2)$.  The matching
coefficients $C_{\text{prod}}$ of the production operators are
computed from the on-shell amplitude for the processes $pp'\to \tilde
s_1\tilde s_2$ at the desired accuracy, as sketched for LO in
Figure~\ref{fig:match-res}.  The coefficients $C_{4p}$ are calculated
by expanding the forward-scattering amplitude in the hard momentum
region, where all loop momenta are far off-shell.\footnote{Explicit
  calculations have been performed for $W$-pair
  production~\cite{Beneke:2007zg} and top-pair
  production~\cite{Beneke:2010mp,Penin:2011gg} at electron-positron
  colliders.}  The leading non-resonant contribution to the
processes~\eqref{eq:example-process} arises from the squared tree
amplitudes of the processes $pp'\to \tilde s_1\tilde s_2' p_2$ and
$pp'\to \tilde s_1' p_1 \tilde s_2$ as sketched in
Figure~\ref{fig:match-nonres} for the example of squark-antisquark
production and decay.
The EFT calculation of the forward-scattering amplitude to order $\alpha_s$ and $\delta$ includes soft one-loop corrections, the one-loop corrections to $C_{\text{prod}}$ as well as kinematic corrections due to subleading kinetic Lagrangian terms.
\subsection{Soft and Coulomb resummation for the resonant contributions}
\label{sec:resum}
The resummation of soft logarithms and Coulomb corrections has been derived
 in~\cite{Beneke:2010da}  by establishing a factorization of the doubly
 resonant contribution to the partonic cross section into
 hard, soft and non-relativistic  (potential) matrix elements,
\begin{equation}
\label{eq:fact}
  \hat\sigma^{\text{res}}_{pp'}(\hat s,\mu)
= \sum_{R_\alpha}H^{R_\alpha}_{pp'}(m_{\tilde q},m_{\tilde g},\mu)
\;\int d \omega\;
J_{R_\alpha}(E+i\,\bar\Gamma-\frac{\omega}{2})\,
W^{R_\alpha}(\omega,\mu)\, .
\end{equation}
Here we have defined the average width
\begin{equation}
\bar\Gamma=\frac{1}{2}(\Gamma_{\tilde s_1}+\Gamma_{\tilde s_2})
\end{equation}
and $R_\alpha$ are the irreducible
colour representations in the decomposition of the product of the
colour representations of the sparticles $\tilde s_1$ and $\tilde
s_2$. 
In analogy to the zero-width treatment in~\cite{Falgari:2012hx}, as a default we use the expression
\begin{equation}
\label{eq:e-beta}
E=\bar m\left(1-\frac{4\bar m^2}{\hat s}\right)
\end{equation} 
for the energy of the squark or gluino pair 
which coincides with the non-relativistic expression $E=\bar m\beta^2$ for $\hat s>4 \bar m^2$. Therefore we refer to this treatment as the `$\beta$-implementation'. In order to estimate ambiguities of the threshold approximation we will also use an `$E$-implementation' defined by $E=\sqrt{\hat s}-2\bar m$ that agrees with~\eqref{eq:e-beta} near threshold.

 The hard functions $H^{R_\alpha}_{pp'}$ are related to the
square of the matching coefficients $ C_{\text{prod}}$. At LO they are
proportional to the Born cross section at threshold.\footnote{As
  in~\cite{Falgari:2012hx}, in our
  numerical results we compute the hard function from the full
  Born cross section instead of its threshold limit.}  The soft functions $W^{R_\alpha}$ are matrix
elements of soft-gluon Wilson lines and implement the eikonal
approximation.  Precise definitions can be found
in~\cite{Beneke:2010da}. Threshold resummation of soft logarithms can
be performed using renormalization group equations for the hard and
soft function~\cite{Becher:2009kw,Beneke:2009rj,Czakon:2009zw} that
can be solved in Mellin-moment or momentum
space~\cite{Becher:2006nr}. The relevant anomalous dimensions 
 for all squark and gluino production processes up to NNLL
accuracy have been collected in~\cite{Beneke:2009rj,Beneke:2010da}.

The potential function $J_{R_\alpha}$ is an expectation value of the
non-relativistic fields $\psi^{(0)}$, that have been decoupled from the soft gluons
by a field redefinition and interact only through the exchange of
Coulomb gluons. It is related to the zero-distance Green function of
the non-relativistic Schr\"odinger equation with a Coulomb potential, $\,G_{\rm C}^{R_\alpha (0)}(0,0;{\cal E})$:
\begin{equation}
J^{R_{\alpha}}({\cal E})
= 2\,\text{Im}\,G_{\rm C}^{R_\alpha (0)}(0,0;{\cal E}) \, ,
\label{eq:potential-green}
 \end{equation}
with the Coulomb
Green function including all-order gluon 
exchange in the 
$\overline{\text{MS}}$-scheme~\cite{Wichmann:1961,Beneke:1999zr}:
\begin{eqnarray}
G_{\rm C}^{R_\alpha(0)}(0,0;{\cal E}) &= & 
-\frac{(2m_{\text{red}})^2}{4\pi} \Bigg\{
    \sqrt{-\frac{{\cal E}}{2m_{\text{red}}}}
    + (-D_{R_\alpha})\alpha_s \bigg[\,
      \frac{1}{2}\ln \bigg(\!
      -\!\frac{8\,m_{\text{red}}{\cal E}}{\mu^2}\bigg)
\nonumber\\
&&    -\,\frac{1}{2}  +\gamma_E
    +\psi\bigg(1-\frac{(-D_{R_\alpha})\alpha_s}{
      2\sqrt{-{\cal E}/ (2m_{\text{red}})}}\bigg)\bigg]\Bigg\}\,.
\label{eq:coulombGF}
\end{eqnarray}
Here $\gamma_E$ is the Euler-Mascheroni constant and
$m_{\text{red}}=m_{\tilde s_1}m_{\tilde s_2}/(m_{\tilde s_1}+m_{\tilde s_2})$.
The coefficients $  D_{R_\alpha}=\frac{1}{2}(C_{R_\alpha}-C_{R_1}-C_{R_2})$ of the Coulomb
potential  depend on the quadratic Casimir operators of the colour representations $R_i$ of the sparticles $\tilde s_i$ and of the irreducible representation $R_{\alpha}$ in the decomposition of $R_1\otimes R_2$.
The trivial ($\alpha_s\to 0$) potential function for finite width obtained by
neglecting the Coulomb corrections is given by
\begin{gather} \label{eq:J0width}
J^{R_\alpha(0)}(E+i\bar\Gamma)=\frac{\sqrt{2}m_{\text{red}}^{3/2}(E^2+\bar\Gamma^2)^{1/4}}{\pi}\frac{E+(E^2+\bar\Gamma^2)^{1/2}}{\sqrt{\left(E+(E^2+\bar\Gamma^2)^{1/2}\right)^2+\bar\Gamma^2}}.
\end{gather}
The use of a complex energy $E+i \bar\Gamma$ takes the finite-width effects in the doubly-resonant cross section into account in leading order in the non-relativistic expansion~\cite{Fadin:1987wz}.

For energies below the production threshold $E<0$ and for $\bar\Gamma=0$, the Coulomb Green function develops a series of bound-state poles at energies
\begin{equation}
\label{eq:bound}
  E_n=-\frac{2m_{\text{red}}\alpha_s^2 D_{R_\alpha}^2}{4 n^2}\,.
\end{equation}
These contributions have been taken into account in the NLL predictions of~\cite{Falgari:2012hx}.
For unstable particles, the complex energy shift in the potential function in~\eqref{eq:fact} leads to a smoothing of the discrete set of bound-state poles~\eqref{eq:bound} into a continuous non-vanishing contribution below the nominal production threshold.


\section{Estimate of contributions to the cross section}
\label{sec:power-counting}

In Section~\ref{sec:frame} we have discussed the treatment of finite-width effects (Section~\ref{sec:eft}) and higher-order soft and Coulomb corrections for the case of stable particles (Section~\ref{sec:resum}).
The resummation of threshold logarithms and Coulomb corrections was performed under the assumption~\eqref{eq:count-as} that a sizable contribution to the total hadronic cross section arises from the region where $\alpha_s\ln\beta\sim  \frac{\alpha_s}{\beta}\sim 1$, leading to a reorganization of the perturbative expansion according to~\eqref{eq:nll-def}. 
For finite decay widths, the replacement $E\to E+i\bar \Gamma$ in the potential function 
 leads to the screening of the threshold singularities, raising the question if the resummation based on the counting~\eqref{eq:count-as} is still justified.
Furthermore, for an unstable particle the size of the non-resonant contribution to the cross section~\eqref{eq:sigma-pole} 
compared to the higher-order corrections to the resonant contributions
has to be addressed.
 In this section we use power-counting arguments to estimate these effects and compare these expectations to explicit results for squark and gluino production.

\subsection{Size of the decay width}
\label{sec:count-gamma}
  To discuss the size of the screening effect and the non-resonant contributions, we consider the decay of the sparticle
$\tilde s$ of mass $M$ into a sparticle $\tilde s'$ of mass $m$ and a
massless parton $p$, mediated by the strong interaction.  On
kinematic grounds, the decay width is of the form
\begin{equation}
\label{eq:decay}
  \Gamma(\tilde s\to \tilde s' p)=  \alpha_s M \left(1-x^2\right)\, f(x)
\end{equation}
for some function $f$ of the mass ratio
\begin{equation}
 x\equiv \frac{m}{M}.
\end{equation}

In the following, we consider the decay
width~\eqref{eq:decay} for two limiting cases:
\begin{enumerate}[a)]
\item Light decay product, $x\to 0$:
  \begin{equation}
    \Gamma/M\sim \alpha_s.
  \end{equation}
In this case, the expansion parameter $\delta$ of the effective theory~\eqref{eq:def-delta} is estimated as
\begin{equation}
\label{eq:scaling-a}
\delta\sim\beta^2\sim \alpha_s .
\end{equation}
\item Heavy decay product, $x\to 1$:
  \begin{equation}
\label{eq:gamma-r1}
    \Gamma/M \sim  \alpha_s \times (1-x^2)^\gamma,
  \end{equation}
where $\gamma=1$ if the function $f(x)$ in~\eqref{eq:decay} is of the order one, but $\gamma=2$ for the case of squark and gluino decay~\eqref{eq:qgdecaywidth}.
In this case, another small scale $\kappa\equiv (1-x^2)^\gamma$  is present, and the hierarchy of scales is given by
\begin{equation}
\label{eq:scaling-b}
  \delta \sim \beta^2 \sim  \alpha_s \kappa.
\end{equation}
In particular, if  for squarks and gluinos the numerical size of the mass
ratio is such that $\kappa=(1-(m/M)^2)^2 \sim 0.1$, i.e. $m/M\gtrsim 0.8$, we count $\delta\sim \alpha_s^2$.
\end{enumerate}

For the phenomenology of unstable strongly-interacting particles, different scenarios have been identified (e.g.~\cite{Hagiwara:2009hq,Kats:2009bv}). 
For very narrow particles, bound states can form that subsequently decay into di-photon or di-jet final states and therefore lead to a very different phenomenology than the missing energy signatures usually employed in squark and  gluino searches. In the MSSM this case is only relevant for gluinonium and stoponium production and will not be considered further here.
For a sparticle decay width that is larger than the bound-state decay widths into di-photons or di-jets, but much smaller than the energy $E_1$ of the first bound-state~\eqref{eq:bound}, a few bound state-peaks can form, but the decay is dominated by the underlying sparticle decay, while for $\bar\Gamma\lesssim E_1$, bound-state effects enhance the cross section near threshold, but the different peaks might not be separated.  Since $E_1\sim\alpha_s^2$, bound-state peaks might be visible in the invariant mass-distribution below threshold  for narrow sparticles in the category b), as we will study in Section~\ref{sec:invmass}.  For $\bar\Gamma> E_1$ the bound-state effects are washed out, which we expect to be the case for broader sparticles in category a).

We will now estimate the size of the Coulomb and soft corrections as well as the non-resonant contributions for these two limiting cases.

\subsection{Coulomb corrections}

\begin{figure}[t]
\centering
\includegraphics[width=0.49 \linewidth]{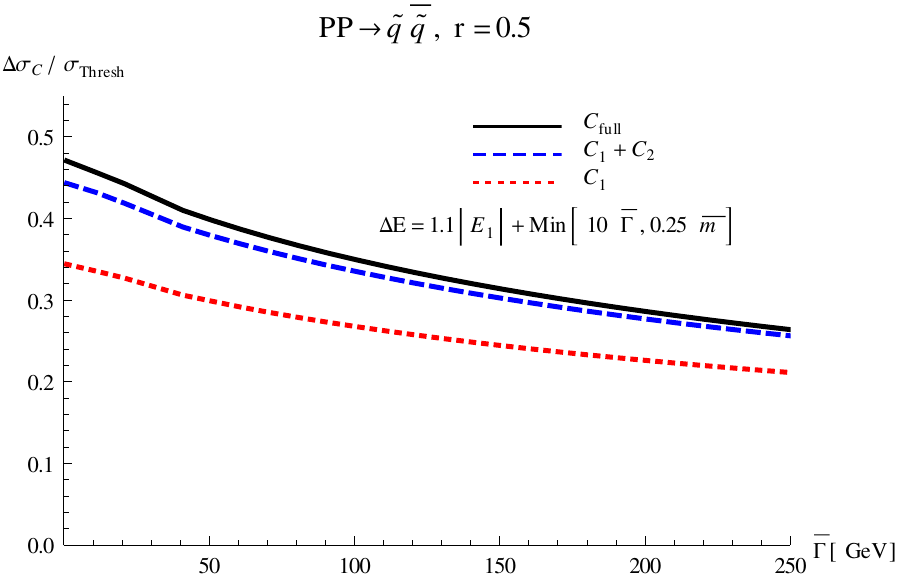}
\includegraphics[width=0.49 \linewidth]{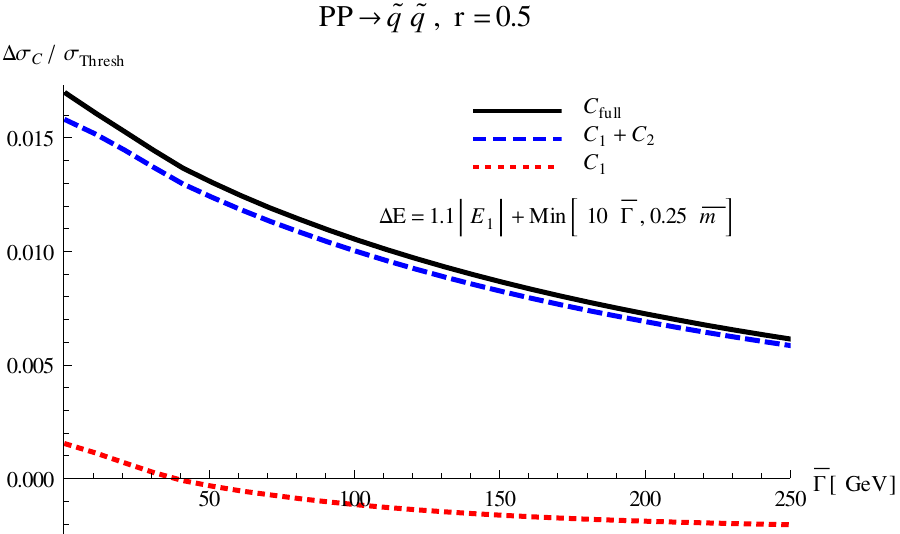}\\
\includegraphics[width=0.49 \linewidth]{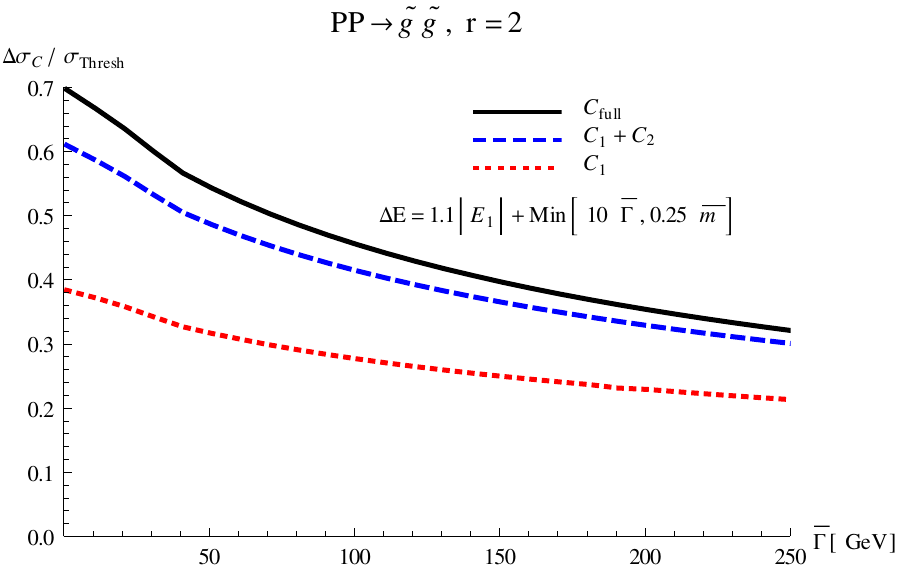}
\includegraphics[width=0.49 \linewidth]{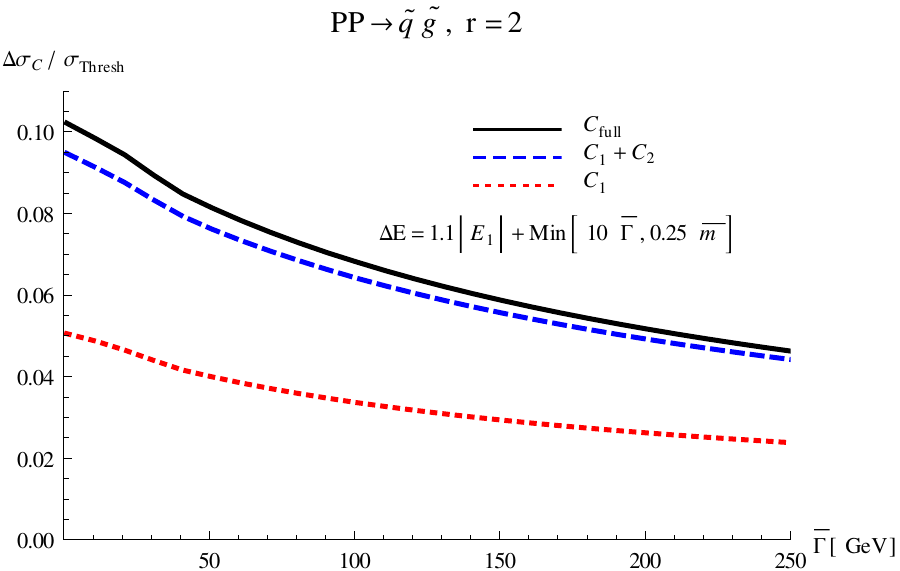}
\caption{Size of the first Coulomb correction (red, dotted), first and second Coulomb corrections (blue, dashed) and resummed Coulomb corrections (black, solid) relative to the leading cross section in the effective theory as a function of $\bar{\Gamma}$ and for $\bar{m}=1500\, \text{GeV}$. $r$ is defined as $r=m_{\tilde{g}}/m_{\tilde{q}}$.  For technical reasons the potential function is set to zero for
 $E<\Delta E$ as defined in \eqref{eq:DeltaEdef}.}
\label{fig:coulomb}
\end{figure}

In the two scenarios for the decay width, the $n$-th Coulomb correction is of the order
\begin{equation} \label{eq:scal_coul}
  \left(\frac{\alpha_s}{\beta}\right)^n
  \sim  \left(\frac{\alpha_s^2}{\delta}\right)^{n/2}
  \sim
  \begin{cases}
 \delta^{n/2}\sim \alpha_s^{n/2},& \text{case a)}\\
  \left(\frac{\alpha_s}{\kappa}\right)^{n/2},& \text{case b)}
 \end{cases}
\end{equation}
Therefore by power counting, Coulomb resummation is not necessary for
case a), i.e. for particles with decay widths of the order of $
\Gamma/M\sim\alpha_s\sim 10\%$.  This exemplifies the screening of the
threshold corrections by the finite lifetime.  However, to reach NLO
accuracy in $\delta\sim\alpha_s$ the second Coulomb correction has to
be included. This is analogous to the case of $W$-pair production
considered in~\cite{Beneke:2007zg} (up to the replacement of $\alpha$
by $\alpha_s$).  In the opposite limit of narrow particles where
numerically $\frac{\Gamma}{M}\sim\alpha_s^2\sim 1\%$ (case b),
$\kappa\sim\alpha_s$ so the screening of the Coulomb corrections is
not effective and Coulomb resummation should be performed as in the
stable case.\footnote{This is analogous to the well-known case of
  top-quark production at linear colliders, where the small decay
  width of the top is due to the electroweak nature of the decay,
  $\Gamma_t/m_t\sim \alpha_{\text{EW}}\sim \alpha_s^2$.} Since in these
formal counting arguments the numerical prefactors such as the
strength of the Coulomb potential are not taken into account, the
actual relevance of the Coulomb corrections has to be studied on a
process-by-process basis. In Figure~\ref{fig:coulomb} we show the
contribution of the first ($C_1$, red dotted line) and second ($C_2$, blue dashed line) Coulomb corrections, as well as
the resummed Coulomb corrections ($C_\text{full}$, solid black line), to the total hadronic production
cross sections as a function of the decay width for all four squark
and gluino production processes.  The Coulomb corrections are
normalized by the approximation $\sigma^{(0)}_{\text{Thresh}}$ that is
defined as the effective theory cross section~\eqref{eq:fact} without
soft-gluon resummation and using only the $\alpha_s^0$ term of the
potential function~\eqref{eq:J0width}.
One sees that for larger widths the
  Coulomb corrections are increasingly saturated by the first two
  Coulomb corrections, while the second Coulomb correction always
  gives a non-negligible contribution, in agreement with the general
  discussion above. However, the total contribution of the Coulomb
  corrections can be sizable, as for gluino-pair production, or
  numerically small, as for squark-squark production.

\subsection{Soft corrections}
The parametric scaling of soft corrections in the scenarios a) and b) considered in Section \ref{sec:count-gamma} 
is given by
\begin{equation} \label{scal_soft}
  \ln \beta
  \sim \ln \sqrt{\delta}
  \sim
  \begin{cases}
 \frac{1}{2} \ln \alpha_s,& \text{case a)}\\
  \frac{1}{2} (\ln \alpha_s+ \ln \kappa) \sim \ln \alpha_s,& \text{case b)}
 \end{cases}
\end{equation}
where in case b) we have again considered the case where numerically $\kappa \sim \alpha_s$. It is thus not obvious at a first glance 
in which scenarios resummation of soft-gluon correction is necessary, if at all, since for $\alpha_s \sim 0.1$ one has 
$\alpha_s|\ln \alpha_s| \sim 0.23$. Clearly one would expect resummation effects
to be more important for case b), where the screening effect due to a finite width is less effective. However the parametric
scaling in the two scenarios is formally the same and equal, up to constant prefactors, to $\sim \ln \alpha_s$. As already
stressed for the case of Coulomb corrections, the scaling argument which leads to \eqref{scal_soft} does not 
take into account the coefficients of the logs, which can be numerically large. This is particularly true for gluon-initiated 
squark-antisquark production and for squark-gluino and gluino-gluino production, where the Casimir invariants for the adjoint 
and higher colour representations appear in the prefactors. Once again, a meaningful assessment of the importance of 
resummation can only be done on a process-by-process basis by a numerical analysis of different fixed-order contributions.     

\begin{figure}[t]
\centering
\includegraphics[width=0.49 \linewidth]{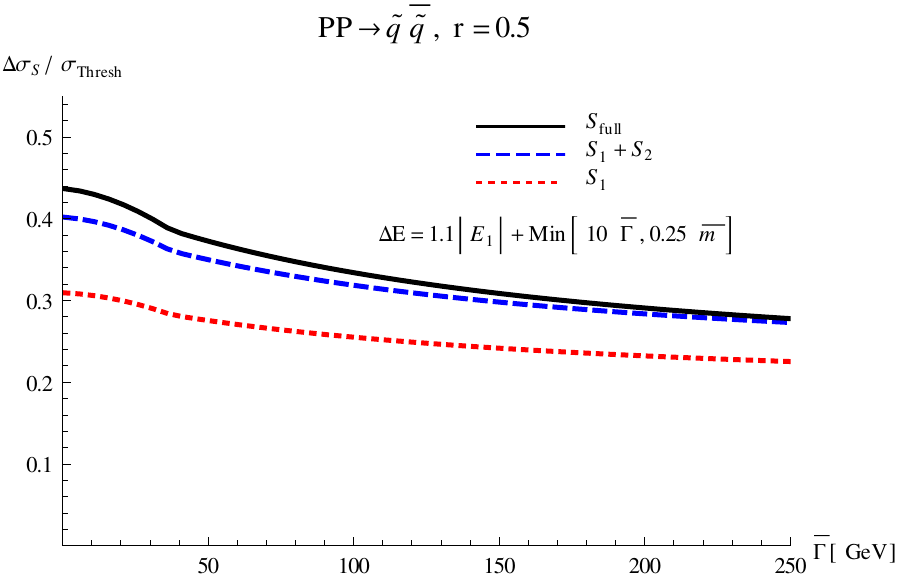}
\includegraphics[width=0.49 \linewidth]{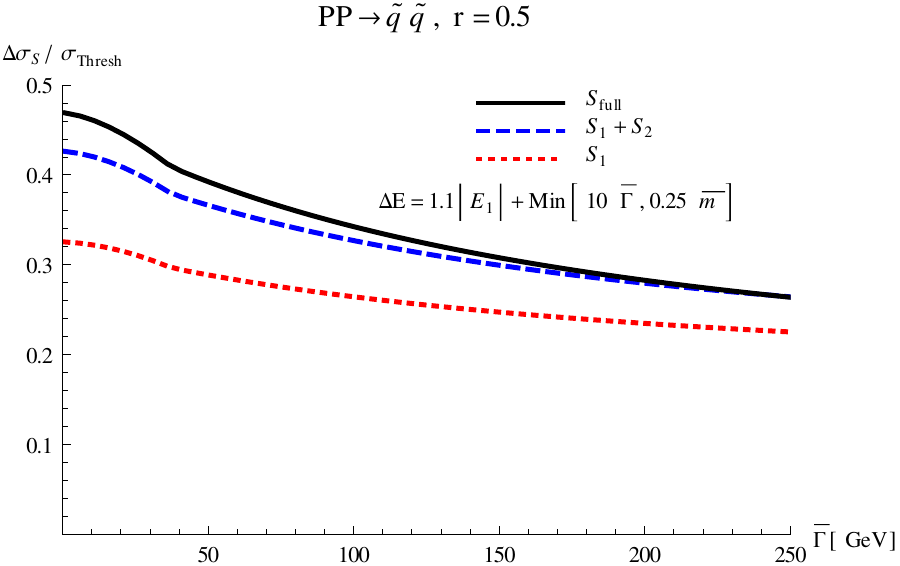}\\
\includegraphics[width=0.49 \linewidth]{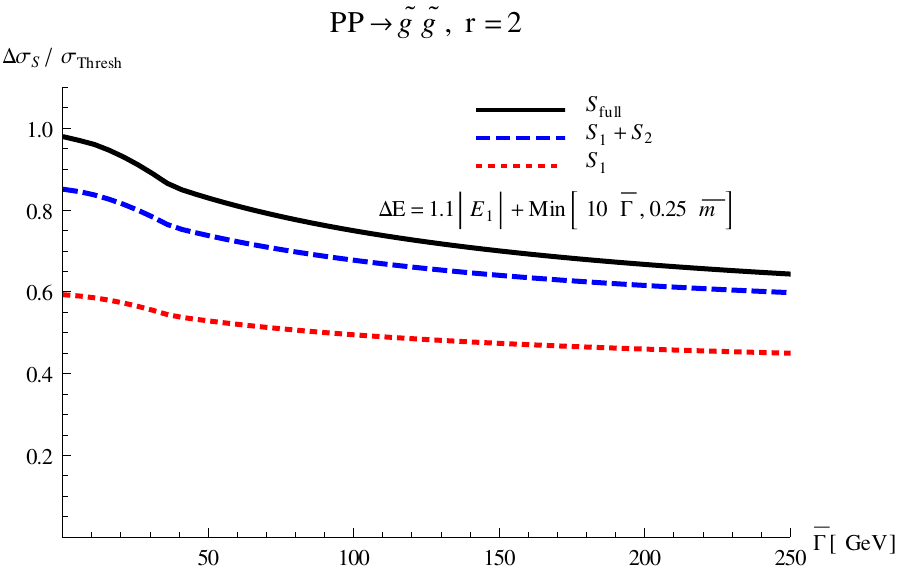}
\includegraphics[width=0.49 \linewidth]{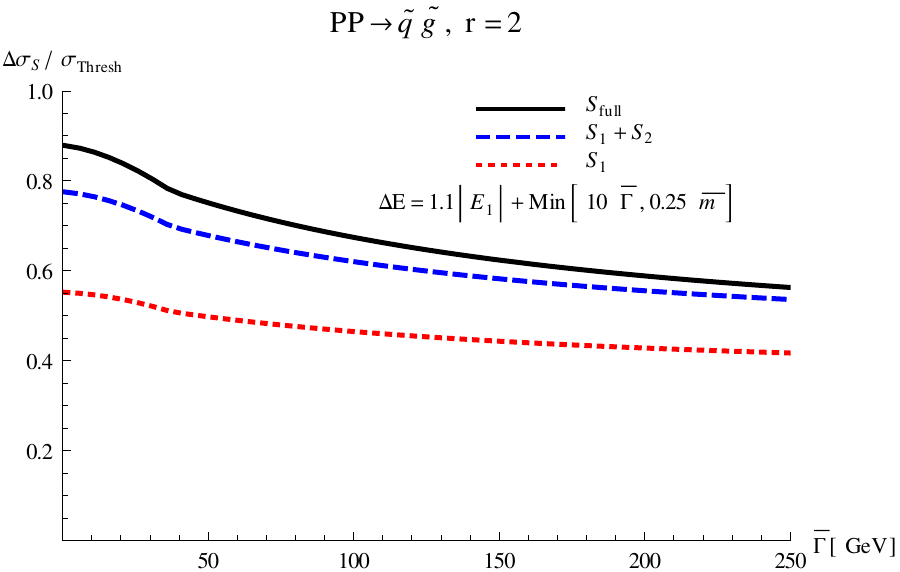}
\caption{Size of the NLL soft  corrections expanded to NLO (red, dotted), up to NNLO (blue, dashed) and the resummed NLL corrections (black, solid) relative to the leading cross section in the effective theory as a function of $\bar{\Gamma}$
 and for $\bar{m}=1500\, \text{GeV}$. The remaining definitions are as in Figure~\ref{fig:coulomb}.
}
\label{fig:soft}
\end{figure}

Figure \ref{fig:soft} shows the effect of NLL soft resummation
($S_\text{full}$, solid black line) on the cross section of the four
SUSY-production processes considered in this work and the
contributions of the ${\cal O}(\alpha_s)$ ($S_1$, red dotted line) and
${\cal O}(\alpha^2_s)$ ($S_2$, blue dashed line) terms obtained from
the expansion of the resummed cross section, as a function of the
width $\bar{\Gamma}$. As in Figure \ref{fig:coulomb} the cross sections
are normalized to the leading threshold approximation.  It can be seen
that for small widths of order $\bar{\Gamma}/\bar m \sim \alpha_s^2$
resummation is numerically important for all processes (and more so
for those where large colour charges are involved), with NNLO
corrections of the order of $10-25\%$ of the LO cross section and
terms beyond NNLO as large as $\sim 10\%$ for squark-gluino and
gluino-gluino production. One can therefore argue that in scenario b)
it makes sense to keep the zero-width scaling $\alpha_s \ln \beta \sim
1$ for soft logarithms, which leads to the representation of the cross
section given in \eqref{eq:nll-def}.

At larger width ($\bar{\Gamma}/\bar m \gtrsim \alpha_s$) the bulk of NLL corrections is accounted for by the ${\cal O}(\alpha_s)$ terms, which correspond to a $30-50\%$ 
contribution to the Born result. However NNLO terms still represent a correction of order $5-15\%$ depending on the process considered.
Consistent with the stronger screening effect expected for larger widths, higher-order soft corrections beyond NNLO are small, $\lesssim 5\%$ for gluino-gluino 
and squark-gluino production and below $1\%$ for squark-antisquark and squark-squark production. We thus conclude that in case a) the relevance of soft resummation
is not immediately clear, though the inclusion of higher-order corrections at NNLO might still be necessary to achieve an accuracy of order $\delta\sim \alpha_s$.
It is interesting to note that for large widths the fixed-order one-loop soft correction is numerically about $\sqrt{\alpha_s}\sim 30\%$ of the tree-level threshold result,
and the two-loop soft terms of order $(\sqrt{\alpha_s})^2 \sim 10\%$.

To summarize, in the two limiting scenarios one can give the following parametric representation of the cross section:    
\begin{enumerate}[a)]
\item Adopting for the soft logarithms the (approximate) scaling $\alpha_s \ln^{2} \beta  \sim \sqrt\delta\sim \sqrt{\alpha_s}$, which is motivated by the numerical results
in Figure \ref{fig:soft}, and using~\eqref{eq:scal_coul}, the cross section can be 
represented as
\begin{equation}\label{eq:syst}
 \hat{\sigma}_{p p'} \propto \,\hat\sigma^{(0)}\,\times
  \begin{cases}
 1 & \text{LO}\\
  \frac{\alpha_s}{\beta}, \alpha_s \ln^{2} \beta & \text{$\mbox{N}^{1/2}$LO}\\
   \alpha_s, \left(\frac{\alpha_s}{\beta}\right)^2,
(\alpha_s \ln^2 \beta)^2, (\alpha_s\ln\beta)^2/\beta, \alpha_s \ln \beta,  \beta^2 &  \text{NLO} \\
\ldots
 \end{cases}
\end{equation}
where the expansion in half-integer powers of $\delta$ is similar to the case of $W$-pair production at a linear collider~\cite{Beneke:2007zg}.
There are no terms linear in $\beta$ because they are known to average out to zero for the total cross section.
\item For the case $\kappa\sim \alpha_s$, the counting is identical to the stable case.
\end{enumerate}
While according to \eqref{eq:syst} resummation is parametrically not necessary for $\bar{\Gamma}/\bar m \gtrsim \alpha_s$, in practice
it is not always clear for which numerical value of $\bar{\Gamma}$ one can switch from the prescription \eqref{eq:nll-def} to \eqref{eq:syst}.
This is particularly true for the transition region where $\alpha_s^2 \lesssim \bar{\Gamma}/\bar m \lesssim \alpha_s$. For this reason in Section
\ref{sec:application} we will use the NLL implementation of \eqref{eq:nll-def} for arbitrary values of the width. This choice has the advantage 
of including, in both limits of a small and large width, all the terms relevant to achieving an accuracy of $\sim \delta$, with the exception
of the $\beta^2$ contribution appearing at NLO in the counting \eqref{eq:syst}. These are however correctly taken into account by matching
the effective-theory result to the exact Born result including the leading finite-width and power suppressed effects, as explained in Section 
\ref{sec:nonres}. 

\subsection{Non-resonant corrections}
\label{sec:nonres}
We now turn to an estimate of the size of non-resonant corrections neglected  in present higher-order calculations that treat squark and gluinos as stable. 
The leading non-resonant contributions as well as subleading kinetic corrections to the double-resonant cross section are both included in a calculation of the full Born cross section of a  process of the form~\eqref{eq:example-process} using  multi-leg Monte Carlo programs~\cite{Hagiwara:2005wg} so the strict EFT treatment is not necessary. Also the non-resonant diagrams are in general well-defined only with kinematic cuts, as mentioned above. 
A study of the full non-resonant effects for all squark and gluino production processes would therefore require a dedicated study taking the realistic selection cuts used by the LHC experiments into account, which is beyond the scope of this work.
We will therefore limit ourselves to an estimate of the unknown
radiative corrections to the non-resonant contributions that cannot
easily be computed in a Monte Carlo program. These are given by
higher-order corrections to the diagrams shown in
Figure~\ref{fig:match-nonres}. The computation of these corrections is
currently an ongoing effort for top-pair production at a linear
collider~\cite{Hoang:2010gu,Beneke:2010mp,Penin:2011gg,RuizFemenia:2012ma}
and the computation for the case of squarks and gluinos is far beyond
the scope of the present work. These corrections would also be included in a complete fixed-order NLO calculation of the process~\eqref{eq:example-process}.

We consider again the two cases introduced in Section~\ref{sec:count-gamma}:
\begin{enumerate}[a)]
\item 
Since in the calculation of the matching coefficients of the four-parton operator all the momenta are large compared to the scale $\delta$, there is no enhancement of these corrections by resonant propagators and one expects them to be of the order~\cite{Beneke:2007zg}  
\begin{equation}
\label{eq:nr-est-a}
   \hat\sigma_{pp'}^{\text{non-res}}\sim \alpha_s^3 \sim 
   \hat\sigma_{pp'}^{\text{res}} \times \frac{\alpha_s}{\beta}.
\end{equation}
For the scaling~\eqref{eq:scaling-a} appropriate for $\Gamma/M\sim 10\%$, 
the leading non-resonant corrections are therefore of the order
$\delta^{1/2}\sim \sqrt{\Gamma/M}$, as the first Coulomb
correction.
The
unknown radiative corrections to the non-resonant contributions~\eqref{eq:nr-est-a} therefore are of the order $\alpha_s
\sqrt{\Gamma/M}\sim\alpha_s^{3/2}$ relative to the leading resonant
cross section,  and therefore  beyond NLO accuracy.

\item For a small mass hierarchy leading to the
  scaling~\eqref{eq:scaling-b} the non-resonant contributions can be
  further expanded according to $\beta\ll \sqrt\rho\ll 1$, with
  $\rho=1-m/M$. For the case of top-quark production the leading
  non-resonant corrections were found to be of the
  order~\cite{Penin:2011gg}
\begin{equation}
  \hat\sigma_{pp'}^{\text{non-res}} \sim \hat\sigma_{pp'}^{\text{res}} \times \sqrt{\frac{\Gamma}{M \rho}}.
\end{equation}
Using~\eqref{eq:gamma-r1} we have
\begin{equation}
  \frac{\Gamma}{M \rho}\sim \frac{\alpha_s (1-x^2)^\gamma}{(1-x)}
  \sim 2\alpha_s\kappa^{(\gamma-1)/\gamma}.
\end{equation}
For squark and gluino decay $\gamma=2$,  so counting $\kappa\sim\alpha_s$ we find  that the suppression of the non-resonant corrections is of the order of $(\frac{\Gamma}{M}\frac{1}{\alpha_s^{1/2}})^{1/2}\sim (\Gamma/M)^{3/8}$ compared to the leading resonant term.
 The unknown radiative  non-resonant corrections are then of the order $(\frac{\Gamma}{M}\alpha_s^{3/2})^{1/2}\sim(\Gamma/M)^{7/8}$, i.e. of a similar magnitude as for the case $x\to 0$. These corrections are beyond NNLL accuracy.
\end{enumerate}
To study the numerical impact of the non-resonant corrections, we consider the following approximations:
\begin{itemize}
\item $\hat{\sigma}^{(0)}_{\text{NWA}}$: The LO cross section calculated using the narrow-width approximation everywhere, i.e. the on-shell production cross section in the $\Gamma\to 0$ limit is multiplied by the decay branching ratios. This approximation is valid for
 $M\gg\Gamma$ and far above threshold \cite{Berdine:2007uv}.  In this work only the decay 
processes (\ref{eq:qgdecay}) are considered and thus $\hat{\sigma}^{(0)}_{\text{NWA}}=\hat{\sigma}^{(0)}_{\text{Tree}}(\Gamma=0)$.
\item $\hat{\sigma}^{(0)}_{\text{Thresh}}$:  The threshold approximated LO cross section in the EFT, using only  the $\alpha_s^0$ term of the potential function~\eqref{eq:J0width}.  The unstable-particle momentum in the 
centre of mass frame is in the potential region (slightly off-shell).
\item $\hat{\sigma}^{(0)}_{\text{Off}}$: The LO cross section obtained using the off-shell doubly-resonant diagrams and fixed-width Breit-Wigner propagators computed with WHIZARD~\cite{Kilian:2007gr} and validated with SHERPA~\cite{Gleisberg:2008ta} when possible. 
This implements tree-level off-shell effects and allows for unrestricted momenta of the unstable particles.\footnote{The selection of doubly-resonant diagrams violates gauge invariance. 
However, we have verified for a selection of different covariant and axial gauges that gauge violation is numerically below $1\%$ for the processes of interest in this work.}
\end{itemize}
\begin{figure}[t]
\centering
\includegraphics[width=0.49 \linewidth]{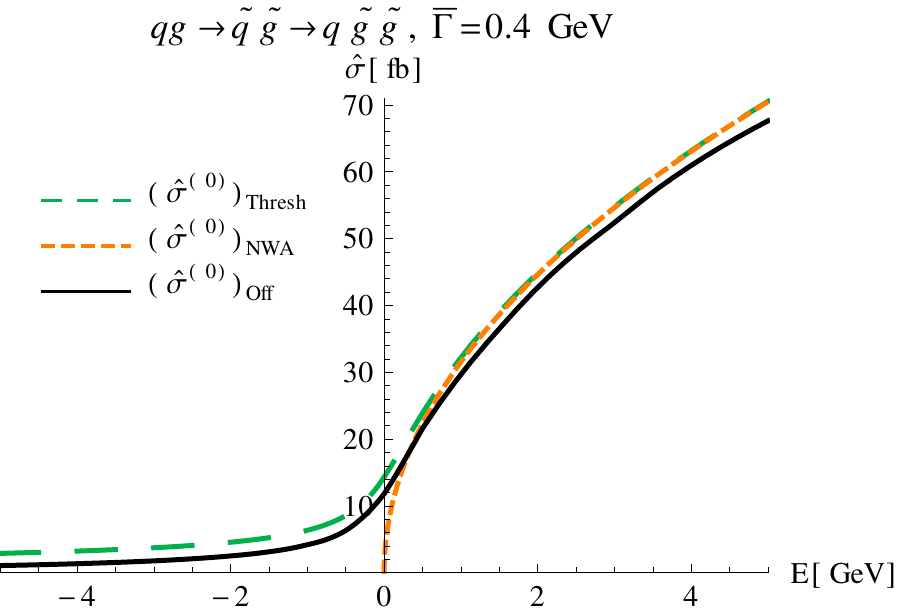}
\includegraphics[width=0.49 \linewidth]{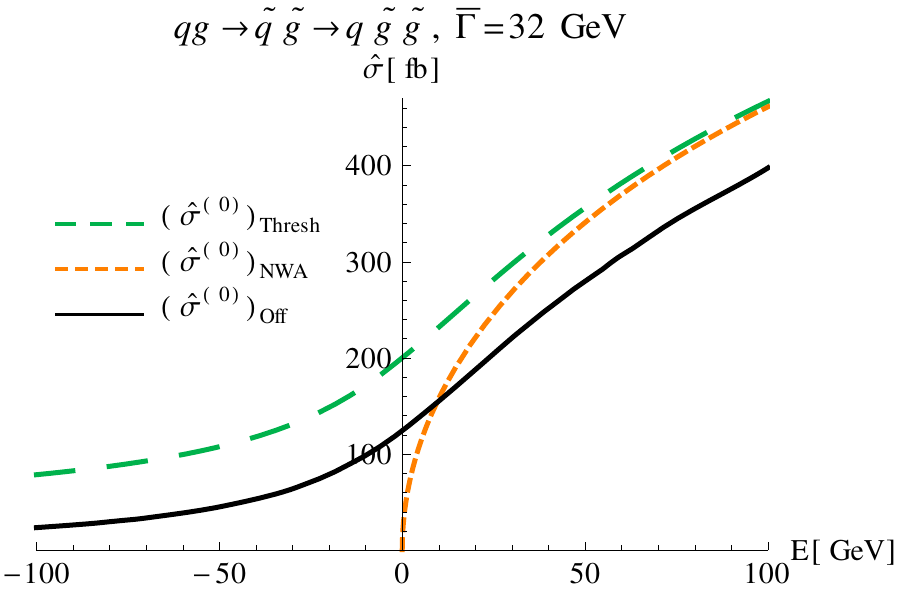}
\caption{The partonic cross section for squark-gluino production and squark decay as a function of the 
energy $E=\bar m\beta^2$ 
at a fixed average mass $\bar m =\frac{1}{2}(m_{\tilde q}+m_{\tilde g})=1500$ GeV. Left: $r=\frac{m_{\tilde g}}{m_{\tilde q}}=0.95,\bar\Gamma=0.4$ GeV. Right: $r=0.5,\bar\Gamma=32$ GeV.} \label{fig:decaypart}
\end{figure}
In Figure \ref{fig:decaypart} the three different approximations of the partonic cross sections are shown for the example of squark-gluino production and subsequent squark decay as  a function of the energy $E=\bar m\beta^2$ for two examples of the decay width. 
The
plots for the other processes have the same behaviour as above and are
not shown here. It can be seen that the EFT approximation
$\hat{\sigma}^{(0)}_{\text{Thresh}}$ agrees with the off-shell results for
small widths, but is systematically shifted for larger decay
width. This is the expected effect of the non-resonant contribution to
the Born cross section, that is of the form $
\hat\sigma_{pp'}^{\text{non-res}}\sim \frac{\bar m^2}{\hat
  s}\text{Im}C_{4p}$, and has  also been observed
in~\cite{Beneke:2007zg,Hoang:2010gu,Beneke:2010mp}.
Note that while we do not include truly singly or non-resonant diagrams in the off-shell approximation, the off-shell momentum configurations of doubly-resonant diagrams contribute to the non-resonant part of the cross section and, in fact, provide the dominant numerical effect in the case of $W$ or top production~\cite{Beneke:2007zg,Beneke:2010mp,Penin:2011gg}. Therefore the shift observed in Figure~\ref{fig:decaypart} provides an estimate of the non-resonant corrections.
For $\bar\Gamma/\bar m=32/1500=0.02$ the nonresonant correction is of the order of $40\%$, which is of the same order of magnitude, but somewhat larger, than the estimate  $(\Gamma/M)^{1/2}$--$(\Gamma/M)^{3/8}= 14\%$--$24\%$. We then estimate that the uncalculated corrections due to higher-order non-resonant effects are of the order of $\lesssim 5\%$.

\section{Application to squark and gluino production at LHC}
\label{sec:application}

In this section we present results for finite-width effects on the
soft and Coulomb NLL corrections to squark and gluino production at
the LHC in the setup discussed in Section~\ref{sec:frame}. 
In Section~\ref{sec:invmass} we give results for the invariant mass distributions, while in Section \ref{sec:results} we discuss our 
implementation  and present numerical results for the total cross sections.

\subsection{Invariant-mass distributions}
\label{sec:invmass}

In order to discuss the finite-width effects on the production cross sections, we first consider the invariant-mass distribution.
 Similarly to  the total cross section~\eqref{eq:fact}, it satisfies a factorization formula near the production threshold~\cite{Beneke:2010da} 
(see also e.g.~\cite{Hagiwara:2009hq})
\begin{equation}
\label{eq:inv-mass}
\frac{d\sigma^{\text{res}}_{NN'}(\hat s,\mu)}{dM_{\tilde s_1\tilde s_2}}
= \sum_{R_\alpha}
J_{R_\alpha}(M_{\tilde s_1\tilde s_2}-2 \bar m+i\bar \Gamma)\!\!\!\!
\sum_{p, p'=q,\bar{q}, g}\!\!\! \!\! 2H^{R_\alpha}_{pp'}(\mu)
\int_{\tau_0}^1\!\!\! d\tau 
\,L_{p p^\prime}(\tau,\mu)\,W^{R_\alpha}(2(\sqrt{\hat s}-M_{\tilde s_1\tilde s_2}),\mu),
\end{equation}
where
$\tau_0=M_{\tilde s_1\tilde s_2}^2/s$.
We have  defined the parton luminosity in terms of the PDFs $f_{p/N}$ for a parton $p$ in a nucleon $N$:
\begin{equation}
\label{eq:lumi}
L_{p p^\prime}(\tau,\mu) = \int_0^1 dx_1
dx_2\,\delta(x_1 x_2 - \tau) \,f_{p/N_1}(x_1,\mu)f_{p^\prime/N_2}(x_2,\mu).\end{equation}
In our numerical results we use the NLL-resummed soft function that can be found in~\cite{Falgari:2012hx}. 
Since we are interested in the invariant-mass spectrum near threshold
we deviate somewhat from the default treatment of the total cross section 
and always express the hard function in terms of the
threshold-approximation of the Born cross section and use a constant
Coulomb scale, defined as solution to the equation $\mu_C =2
\alpha_s(\mu_C) m_r |D_{R_\alpha}|$, and constant soft scales $\mu_s$ that are process-specific and determined in~\cite{Falgari:2012hx}.

Consistent with the treatment of the total cross section discussed in Section~\ref{sec:resum}, as a default we use the expression~\eqref{eq:e-beta} for the energy of the squark or gluino pair, i.e. we write the argument of the soft function $W^{R_\alpha}$ in~\eqref{eq:inv-mass} as
\begin{equation} \label{transf}
\sqrt{\hat s}-M_{\tilde s_1\tilde s_2}\to 
\bar m\left(1-\frac{4\bar m^2}{\hat s }\right) +2\bar m -M_{\tilde s_1\tilde s_2}.
\end{equation}
As for the total cross section, this  will be referred to as the
`$\beta$-implementation'. Note that the soft function vanishes for
negative argument which implies that the lower boundary of the $\tau$
integral is given by
\begin{equation}
  \tau_0=\frac{4\bar m^2}{s}\frac{\bar{m}}{3\bar m-M_{\tilde s_1\tilde s_2}}.
\end{equation}
In order to estimate ambiguities of our treatment, we also introduce
an `$E$-implementation' keeping the unexpanded expression $\sqrt{\hat
  s}-M_{\tilde s_1\tilde s_2}$ in the argument of the soft function. In order not to introduce artificial 
differences in the two implementations due to a different boundary of the $\tau$ integral, we perform the replacement 
$M_{\tilde s_1\tilde s_2}\to M_{\tilde s_1\tilde s_2}'\equiv \bar{m}(3-4\bar{m}^2/M_{\tilde s_1\tilde s_2}^2)$ in the
$E$-implementation and consider the differential cross section with respect 
to $M_{\tilde s_1\tilde s_2}'$.
\begin{figure}[p]
\centering
\includegraphics[width=0.49 \linewidth]{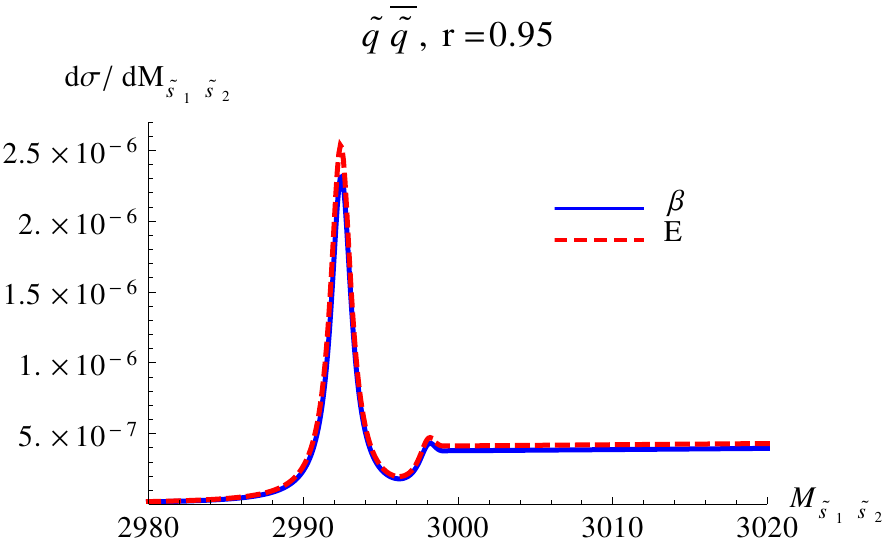}
\includegraphics[width=0.49 \linewidth]{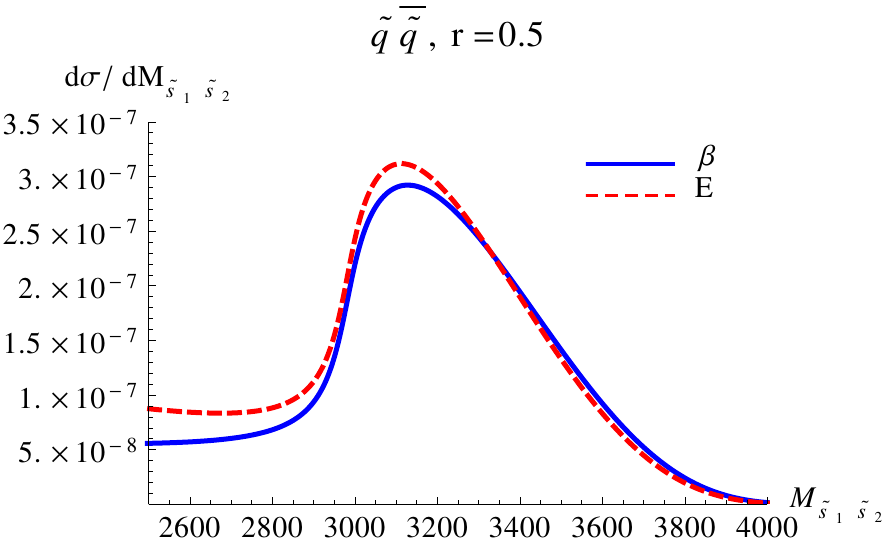}\\
\includegraphics[width=0.49 \linewidth]{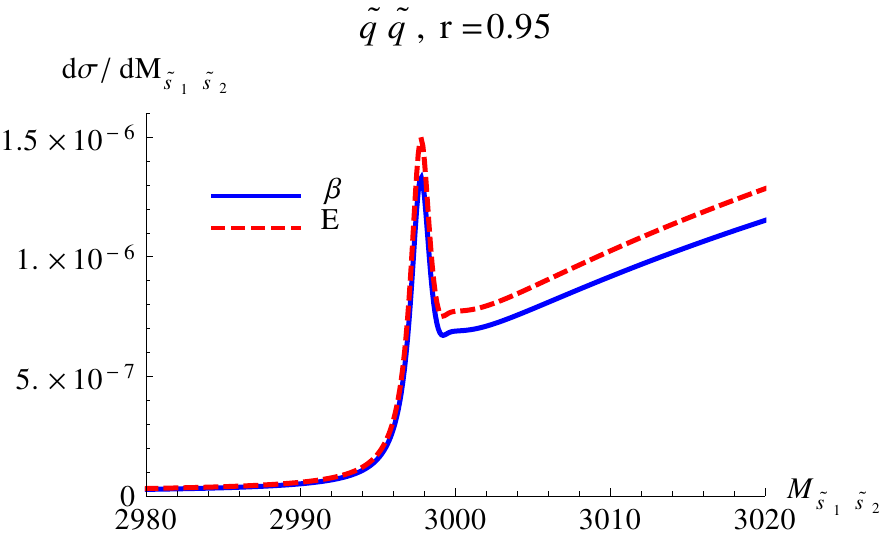}
\includegraphics[width=0.49 \linewidth]{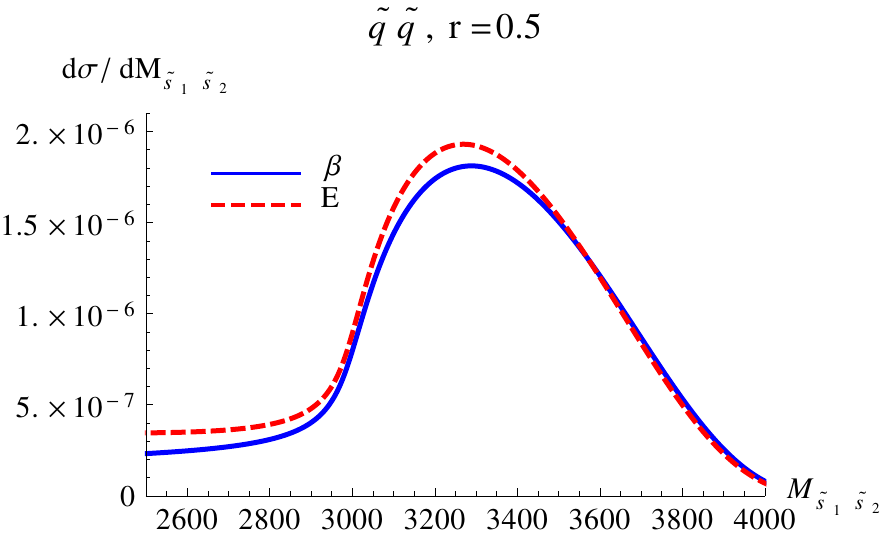}\\
\includegraphics[width=0.49 \linewidth]{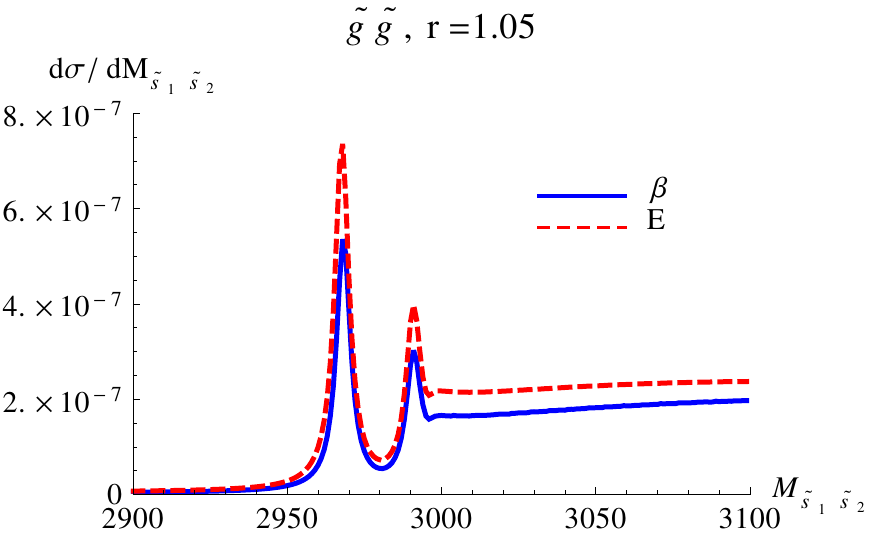}
\includegraphics[width=0.49 \linewidth]{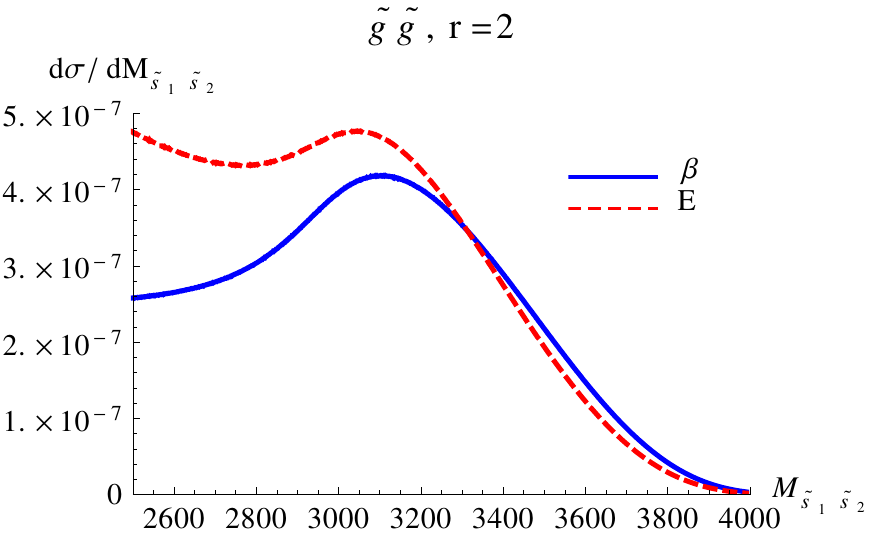}\\
\includegraphics[width=0.49 \linewidth]{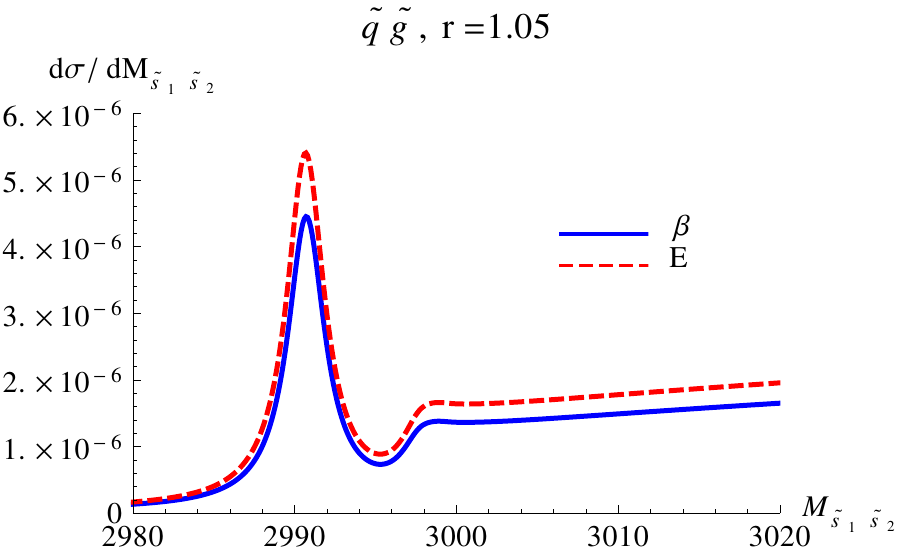}
\includegraphics[width=0.49 \linewidth]{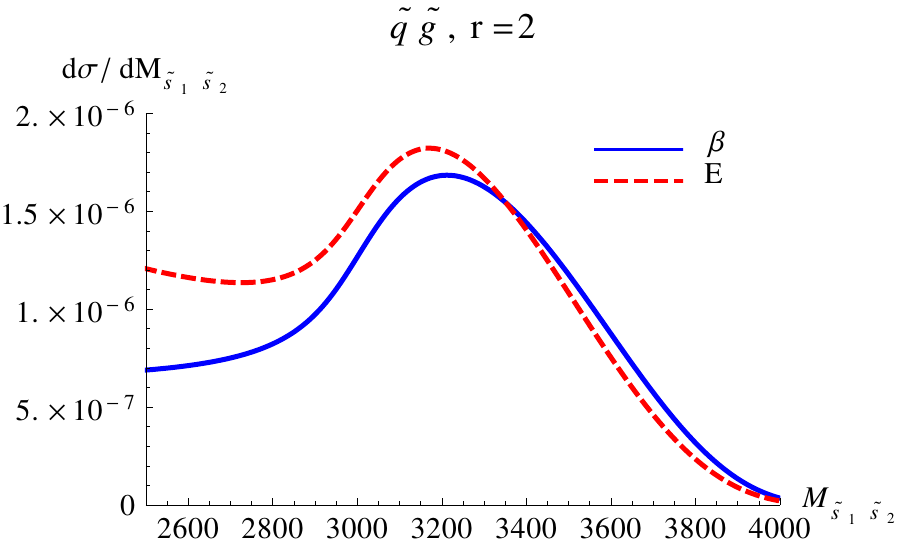}
\caption{Invariant-mass distributions for an average mass $\bar
  m=1500$~GeV of the produced sparticle pair and a centre-of-mass
  energy $\sqrt{s}=8$ TeV for the $\beta$ (blue, solid) and $E$ (red, dashed)
  implementation as defined below~\eqref{transf}. 
Left-hand plots are examples of case b) ($\bar{\Gamma}/\bar m \sim \alpha_s^2$), right-hand plots
case a) ($\bar{\Gamma}/\bar m \sim \alpha_s$).} 
\label{fig:invmass}
\end{figure}

Based on experience with $W$-pair production~\cite{Beneke:2007zg} we do not expect the 
leading resonant approximation to be valid significantly below the nominal threshold, where 
higher-order terms in the non-relativistic expansion and non-resonant contributions become important. 
The differences between the default $\beta$ and the $E$ approximation provide an 
estimate for the size of the higher-order relativistic effects. Note that because of soft radiation (as taken 
into account by the convolution of the soft function with the parton luminosity) also the invariant mass 
distribution directly at threshold is sensitive to larger partonic centre-of-mass energies and the 
associated ambiguities in the threshold expansion.

In Figure \ref{fig:invmass} the invariant-mass distributions for the
four relevant squark and gluino production
processes~\eqref{eq:decayprocesses} are shown for $\bar m=1500$~GeV at
a centre-of-mass energy $\sqrt{s}=8$ TeV for two different mass ratios
corresponding to the cases a) and b) discussed in Section
\ref{sec:count-gamma} . For the case $m_{\tilde g}>m_{\tilde q} $ we
consider $r=m_{\tilde g}/m_{\tilde q}=2$ and $r=1.05$, corresponding
to $\Gamma_{\tilde g}/m_{\tilde g}=12\%$ and $\Gamma_{\tilde g}/m_{\tilde g}=0.2\%$,
respectively.  For the case $m_{\tilde g}<m_{\tilde q} $ we consider
$r=0.5$ and $r=0.95$, corresponding to $\Gamma_{\tilde q}/m_{\tilde q}=3\%$ and
$\Gamma_{\tilde q}/m_{\tilde q}=0.05\%$, respectively.  
As explained in Section~\ref{sec:power-counting}, we will perform treshold resummation also for the case of larger decay widths, although it might not be strictly necessary in this case.

In the small-width case, i.e. the left-hand plots in
Figure~\ref{fig:invmass} where $r=0.95$ or $r=1.05$, the ambiguities in
our prediction as estimated by the difference of the $E$- and $\beta$-implementations are moderate. Depending on the process, the different
regimes mentioned in Section~\ref{sec:count-gamma} are observed: for
the given choice of parameters, a single would-be bound-state peak
appears for squark-squark production, where $\bar \Gamma=0.9\,
\text{GeV}\lesssim E_1=1.5\, \text{GeV}$, while two peaks are visible
in gluino-pair production where $\bar \Gamma=3\, \text{GeV}\ll
E_1=30\, \text{GeV}$. In the remaining two processes the second 
bound-state peak is less pronounced, but still visible.

In the case of larger widths, i.e. the right-hand plots in
Figure~\ref{fig:invmass} with $r=0.5$ or $r=2$, the decay widths are always 
much larger than the bound-state energy, so no peaks below threshold are observed. 
It can be seen that the invariant-mass distributions do not vanish for $M_{\tilde s_1\tilde s_2}$ far below the nominal threshold, 
$M_{\tilde s_1\tilde s_2}=3000$~GeV. This behaviour arises as a combination of the rise of the PDFs for small $x$ and the behaviour of the potential function 
in \ref{eq:inv-mass}, that scales as $J_{D_R}\propto \bar\Gamma/(2\bar m-M_{\tilde s_1\tilde s_2})^{1/2}\to\bar\Gamma/(2\bar m)^{1/2}$ 
for $M_{\tilde s_1\tilde s_2}\to 0$, which follows 
from the LO (trivial) potential function~\eqref{eq:J0width}.
While for the default $\beta$-implementation the distributions approach a flat plateau far below threshold, in the 
$E$-implementation one observes a (clearly unphysical) rise for larger widths for processes involving gluinos.\footnote{Note that this is not an artifact of our use of  
$M_{\tilde s_1\tilde s_2}'$ for the $E$-implementation. In fact, for the original formula~\eqref{eq:inv-mass} an even stronger rise is observed, since smaller $x$ values 
are probed in the parton luminosity due to smaller values of $\tau_0$.} 
In fact, as already pointed out, far below threshold the LO
potential function is not expected to be a good approximation, and the inclusion of non-resonant
contributions and relativistic corrections are needed for a realistic
description of the invariant mass distributions in this region,
similar to what was seen for the partonic cross sections in Figure \ref{fig:decaypart}.

The difference between the $E$- and $\beta$-implementations is amplified
 due to the soft function in \eqref{eq:inv-mass},
$W^{R_\alpha}(\omega)\sim \omega^{-1+2 \eta}$ with $\eta<0$. In the $E$-implementation the
argument of the soft function (left-hand side of Eq. (\ref{transf})) 
approaches zero faster than the corresponding quantity in the $\beta$-implementation (right-hand side of Eq. (\ref{transf}))
as the lower integration boundary $\tau_0$ is approached, and the soft function $W$ diverges faster.\footnote{This is 
further enhanced by the rise of the PDFs at the lower integration
boundary.}  If soft corrections are not taken into account, the soft function
reduces to a delta function, $W^{R_\alpha}(\omega) \propto \delta(\omega)$. In this case we have observed a substantial reduction in the difference between the $E$- and $\beta$-implementation  also for large widths. 

From the results presented in this section we conclude that our approximation is reasonably
accurate for processes with decay widths of the order of
$\bar\Gamma/\bar m\lesssim 5\%$.  As we will argue in Section~\ref{sec:results}, this covers 
the phenomenologically-relevant MSSM parameter-space regions.
Our effective-theory treatment becomes inadequate for larger decay widths, so our results in 
this regime should be considered as indicative only.

\subsection{Total cross sections}
\label{sec:results}

In order to compute the total cross section for a finite decay width,
in principle the $\omega$-integral in~\eqref{eq:fact} has to be
performed up to the maximal energy $\omega=2\sqrt{\hat s}$ since
the potential
function $J_{R_\alpha}$ is defined for arbitrary negative values of
the partonic energy $E$.  However, as mentioned above, far below threshold
(i.e. $\omega \gg 2 E$) the leading threshold approximation becomes
insufficient and higher-order and non-resonant contributions become
relevant. This was shown by the difference of the
$\hat\sigma^{(0)}_{\text{thresh}}$ and $\hat\sigma^{(0)}_{\text{Off}}$
curves for large widths in Figure~\ref{fig:decaypart} and by the large
difference of the $\beta$- and $E$-implementations for the
invariant-mass spectrum and the unphysical behaviour of the latter
below threshold, as seen in Figure~\ref{fig:invmass}.  Since the
non-resonant and higher-order non-relativistic corrections for squark
and gluino production are not available beyond LO, we introduce a
cut-off in the potential function:
\begin{equation} \label{eq:Jwidth}
J_R(E+i\bar \Gamma)=2\text{Im}\left[G_R\left(E+i\bar\Gamma\right)\right]\theta\left(E+\Delta E\right), \quad \Delta E>0,
\end{equation}
with $E$ equal to the expression~\eqref{eq:e-beta} as our default implementation.
The parameter  $\Delta E$ cuts off the convolution of the soft and potential function at $\omega=2(E+\Delta E)$. 
In order to capture the bound-state peaks, this cut-off is chosen as 
\begin{equation} \label{eq:DeltaEdef}
\Delta E=1.1|E_1|+\overline{\Delta E},  
\end{equation}
where $E_1$ is the energy of the lowest-lying bound-state peak, see Eq. (\ref{eq:bound}). 
The prefactor $1.1$ multiplying $E_1$ insures that the bound-state contribution is always
fully included.
As a default for $\overline{\Delta E}$ we take
\begin{equation}
  \overline{\Delta E_0}:=\text{Min}[10\, \bar\Gamma,0.25\,\bar m].
\end{equation}
A related cutoff  $\Delta E=|E_1|+3\bar\Gamma$ was used for gluino-pair production in~\cite{Hagiwara:2009hq} where the focus was on narrow gluinos with decay widths at most of the order of $\Gamma_{\tilde g}\sim |E_1|\sim m_{\tilde g}\alpha_s^2$. In the case of larger decay widths, the value of $\Delta\overline E =10\bar\Gamma$ becomes of the order of $m_{\tilde g}$, where we expect the threshold approximation to become invalid. Therefore we have chosen the cutoff  $\overline{\Delta E}\sim 0.25\bar m$ in the case of larger decay widths.
The uncertainty in choosing $\overline{\Delta E}$ is estimated by varying it around its default value:
\begin{gather}
\quad \overline{\Delta E}_0/2\leq\overline{\Delta E}\leq 2\overline{\Delta E}_0. \label{eq:deltaE}
\end{gather}

The hadronic cross sections are obtained by convoluting the partonic cross sections with the parton luminosity~\eqref{eq:lumi}
\begin{equation}
\label{eq:sigma-had}
\sigma_{N_1 N_2 \to \tilde{s} \tilde{s}'X}(s) = \int_{\tau_0}^1 d \tau \sum_{p, p'=q,\bar{q}, g} L_{p p'}(\tau,\mu_f) \hat{\sigma}_{p p'}(\tau s, \mu_f) \, ,
\end{equation} 
where the lower integration boundary depends on the cutoff, 
\begin{equation}
\tau_0(\Delta E)=\frac{4 \bar m^2}{s}\frac{1}{1+\Delta E/\bar m}  .
\end{equation}
In the $E$-implementation, the lower integration boundary is instead
given by $\tau_0(\Delta E)=(2\bar m-\Delta E)^2/s$. In order not to
introduce numerical differences due to different integration boundaries, in the $E$-implementation  we use the same numerical value for $\tau_0$ as 
for the $\beta$-implementation, and adjust the value of $\Delta E$ in the potential function accordingly.\footnote{This is related to the replacement of the 
invariant mass $M_{\tilde s_1\tilde s_2}\to M_{\tilde s_1\tilde s_2}'$ for the $E$-implementation in Section \ref{sec:invmass}.}

While the Born amplitudes for the production and decay processes of
squarks and gluinos can be treated exactly with Monte
Carlo programs~\cite{Hagiwara:2005wg}, the NLO corrections for the
production processes are known only in the narrow-width approximation, as
implemented in Prospino~\cite{Beenakker:1996ed}.  In order to take
these exactly-known contributions into account, we match our
finite-width NLL results to the sum of the Born cross section including finite-width 
effects and the NLO corrections in the narrow-width
approximation.  In practice, we approximate the Born cross section by
$\sigma^{(0)}_{\text{Off}}$ defined in Section~\ref{sec:nonres} and
computed with Whizard.
The total hadronic cross sections at NLO for finite
widths are thus approximated as:
\begin{equation} \label{eq:NLOwidth}
\sigma_{\text{NLO}}(\bar\Gamma)=\sigma^{(0)}_{\text{Off}}(\bar\Gamma)+(\sigma_{\text{NLO,NWA}}-\sigma^{(0)}_{\text{NWA}}).
\end{equation}
Here $\sigma_{\text{NLO,NWA}}$ is the NLO result obtained with Prospino and $\sigma^{(0)}_{\text{NWA}}$ is the leading-order production cross section for stable squarks or gluinos (see Section~\ref{sec:nonres}). 
Analogously to the  narrow-width case \cite{Falgari:2012hx}, the NLL resummed cross section is then matched to $\sigma_{\text{NLO}}(\bar\Gamma)$:
\begin{eqnarray} \label{eq:matchwidth}
\sigma_{\text{NLL}}^{\text{matched}}(\bar\Gamma)
& = & \Delta\sigma_{\text{NLL}}(\bar\Gamma)
+\sigma_{\text{NLO}}(\bar\Gamma), \nonumber\\
\Delta\sigma_{\text{NLL}}(\bar\Gamma) 
& = & \sigma_{\text{NLL}}(\bar\Gamma)
-\sigma_{\text{NLL, LO}}(\bar\Gamma)
-\sigma_{\text{NLL, NLO}}(\bar\Gamma=0).
\end{eqnarray}
Here $\sigma_{\text{NLL, (N)LO}}$ is the expansion of the NLL resummed cross section to (N)LO accuracy. For comparison we will also consider results where 
the decay width is set to zero in the higher-order corrections, i.e.
\begin{equation}
\label{eq:matchzero}
  \sigma_{\text{NLL}}^{\text{matched}}(0)
 = \Delta\sigma_{\text{NLL}}(0)
+\sigma_{\text{NLO}}(\bar\Gamma)\, .
\end{equation}
The result~\eqref{eq:matchwidth} is our best current prediction. It
could be made more realistic by replacing the off-shell approximation
for the tree cross section by a full LO simulation including
acceptance cuts, which we leave for further investigation. Contrary to
the NLO cross section $\sigma_{\text{NLO}}(\bar\Gamma)$, the matched
NLL cross section $\sigma_{\text{NLL}}(\bar\Gamma)$ does contain
finite-width effects at NLO from the resummed Coulomb and soft
corrections.  It is also interesting to consider only the finite-width
corrections beyond NLO. This shows the size of the effects that are
beyond a possible future exact NLO calculation of the
processes~\eqref{eq:decayprocesses}. For this purpose we consider the
alternative matching
\begin{eqnarray} \label{eq:matchwidth2}
\sigma_{\text{NLL}}^{\text{matched,(2)}}(\bar\Gamma)
& = & \Delta\sigma_{\text{NLL}}^{(2)}(\bar\Gamma)
+\sigma_{\text{NLO}}(\bar\Gamma), \nonumber\\
\Delta\sigma_{\text{NLL}}^{(2)}(\bar\Gamma)
& = & \sigma_{\text{NLL}}(\bar\Gamma)
-\sigma_{\text{NLL, LO}}(\bar\Gamma)
-\sigma_{\text{NLL, NLO}}(\bar\Gamma).
\end{eqnarray}

For the results shown in this section we take as a representative
average mass $\bar m=1500$~GeV  and the setting is the LHC at a centre-of-mass energy of $\sqrt{s}=8$ TeV.
We use the $\beta$-implementation as default and the scales are taken to be the same as in the zero-width case \cite{Falgari:2012hx}: 
$\mu_f=\bar m=1500$ GeV, $\mu_h=2\bar m=3000$ GeV and a running soft and Coulomb scale. We note that the soft and Coulomb scales are 
always fixed below threshold. The LO widths in Eq. (\ref{eq:qgdecaywidth}) for the squarks and gluinos are used.

\begin{figure}[t!]
\centering
\includegraphics[width=0.49 \linewidth]{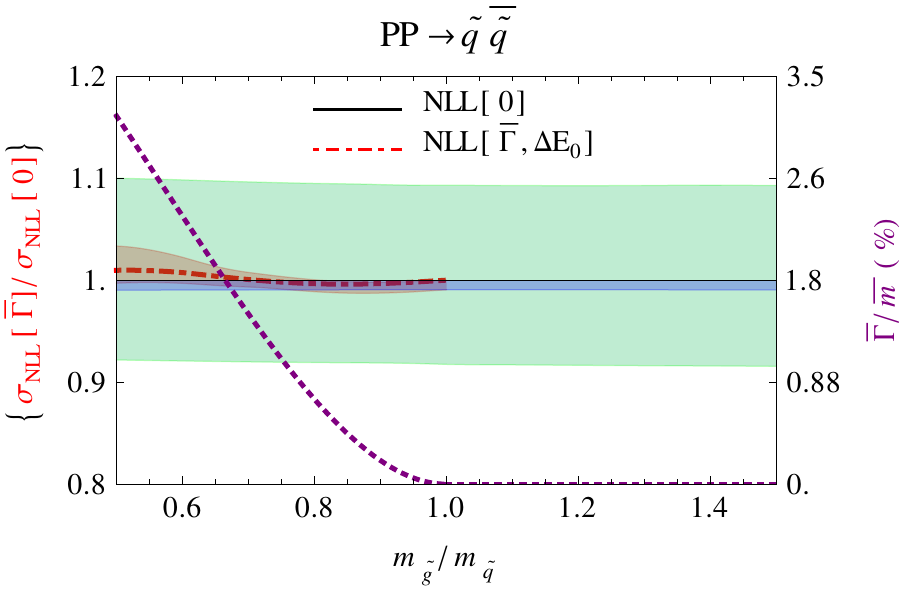}
\includegraphics[width=0.49 \linewidth]{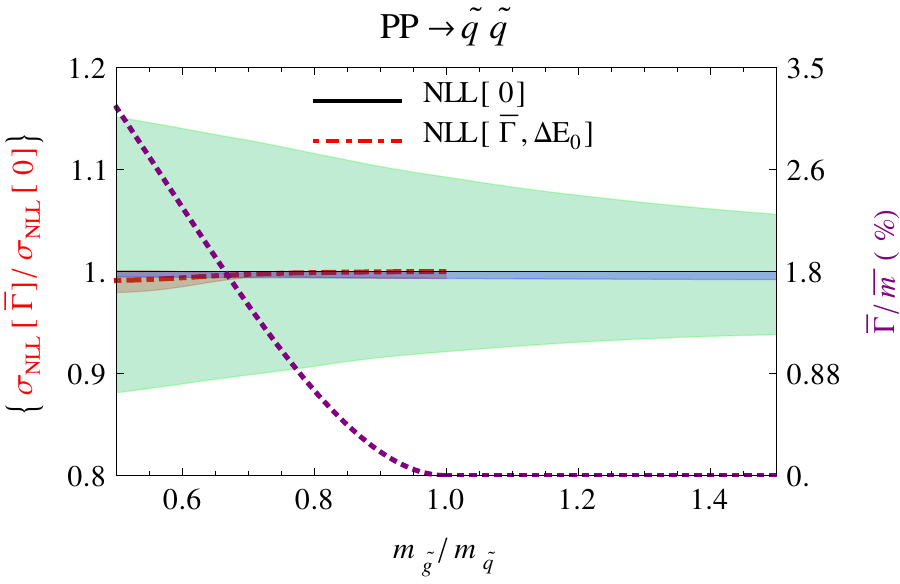}\\
\includegraphics[width=0.49 \linewidth]{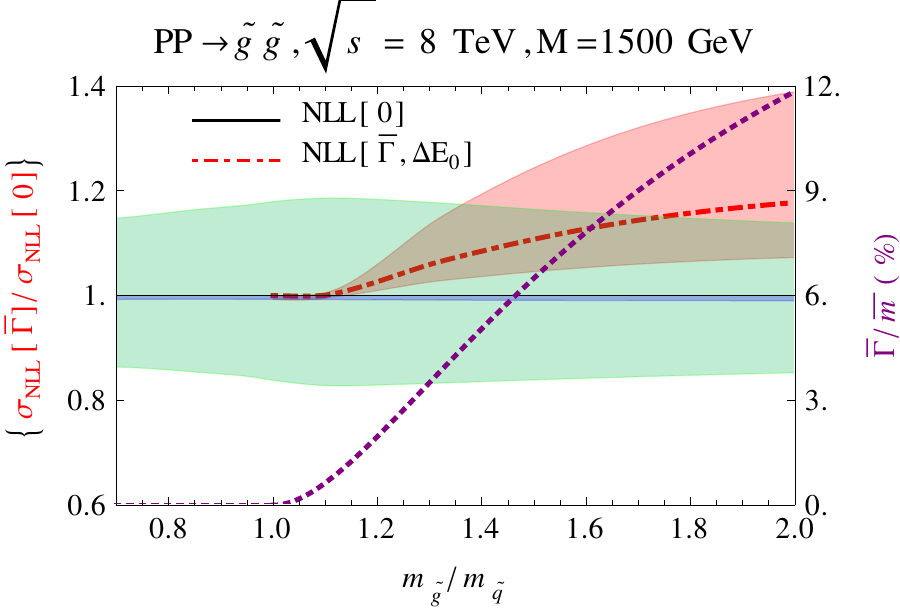}
\includegraphics[width=0.49 \linewidth]{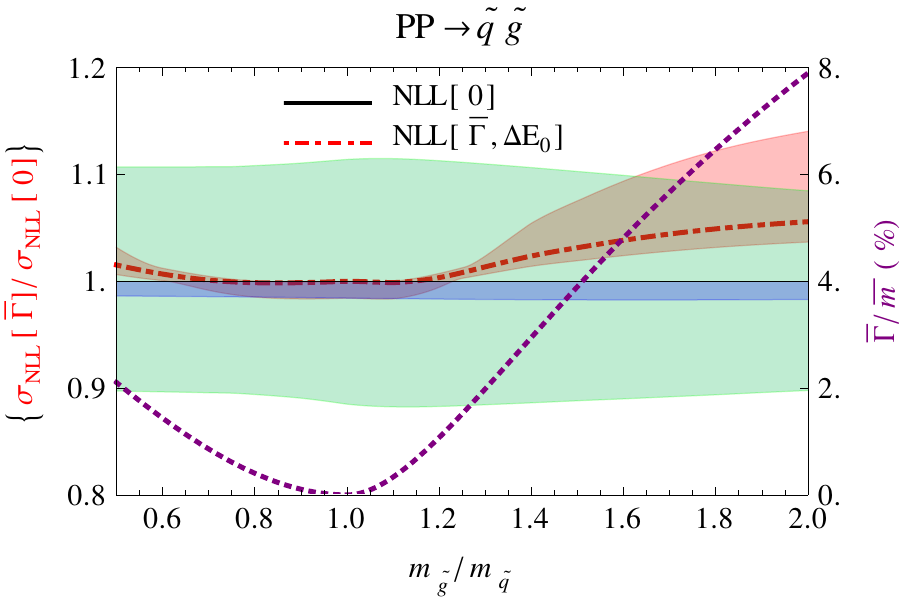}
\caption{The plots show the ratio $\sigma_{\text{NLL}}^{\text{matched}}(\bar\Gamma)/\sigma_{\text{NLL}}^{\text{matched}}(0)$ of the matched finite-width NLL cross section~\eqref{eq:matchwidth}
 to the zero-width approximation~\eqref{eq:matchzero} (red dot-dashed, left-hand vertical axis) for $\bar m=1500$~GeV and $\sqrt{s}=8$ TeV 
and the ratio $\bar\Gamma/\bar m$ in
per-cent (purple dotted, right-hand vertical axis). 
Both quantities are plotted as a function of the gluino-to-squark mass ratio. 
The NLL result is for default $\Delta E=\Delta E_0$.
The red band represents the error from $\overline{\Delta E}$ variation and $E$ vs $\beta$ uncertainty added in quadrature,
the green band the total error for zero width, 
the blue band the $E$ vs $\beta$ uncertainty for the zero-width case.} \label{fig:NLLratio1}
\end{figure}

Figure~\ref{fig:NLLratio1} shows the ratio of the matched NLL cross
section for finite widths~\eqref{eq:matchwidth} to that for the
zero-width case~\eqref{eq:matchzero} (red, dot-dashed) as a function of the gluino-to-squark mass ratio.
The green band gives the total uncertainty for the zero-width NLL result, including ambiguities in soft, hard, factorization and Coulomb-scale choices 
and the difference of the $E$- and $\beta$-implementation, added up in quadrature as
discussed in detail in~\cite{Falgari:2012hx}.
The difference of the $E$- and $\beta$-implementation alone is given by the blue band.
The red band indicates the ambiguities related to our implementation of the finite-width effects, 
as estimated by the $\overline{\Delta E}$ variation as described in Eq. (\ref{eq:deltaE}) 
and the difference of the $E$- and $\beta$-implementations, summed in quadrature. 
The plots also show the ratio of the average decay width to the average mass of the produced particles.

\begin{figure}[t!]
\centering
\includegraphics[width=0.49 \linewidth]{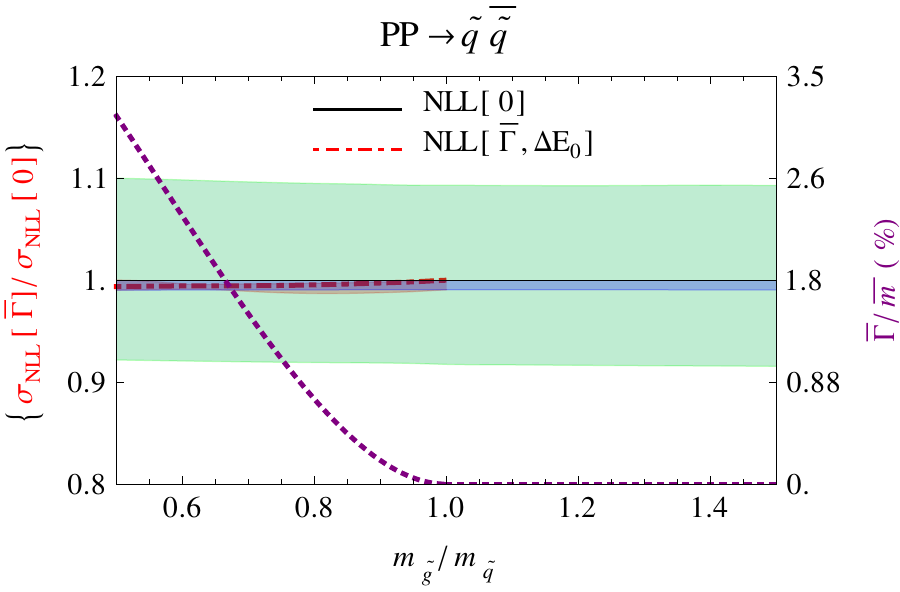}
\includegraphics[width=0.49 \linewidth]{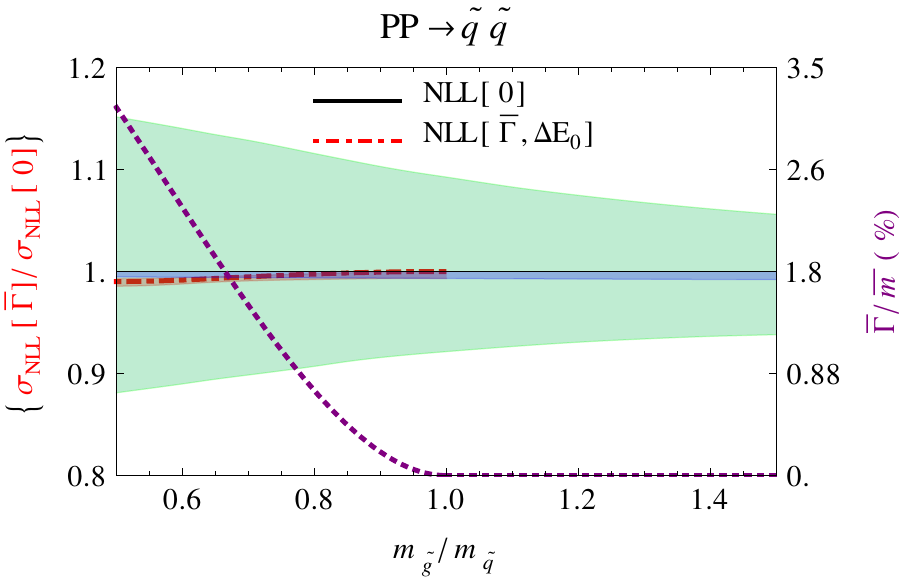}\\
\includegraphics[width=0.49 \linewidth]{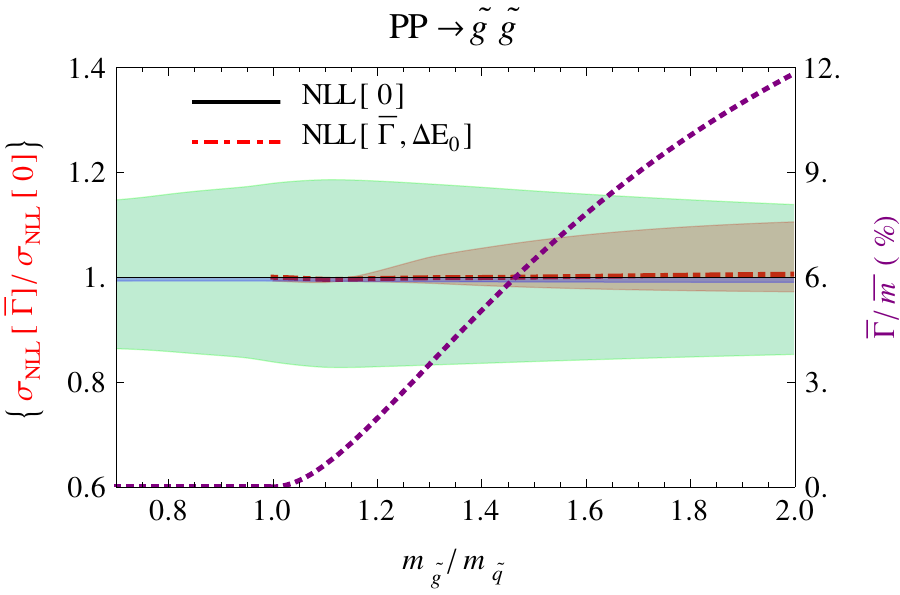}
\includegraphics[width=0.49 \linewidth]{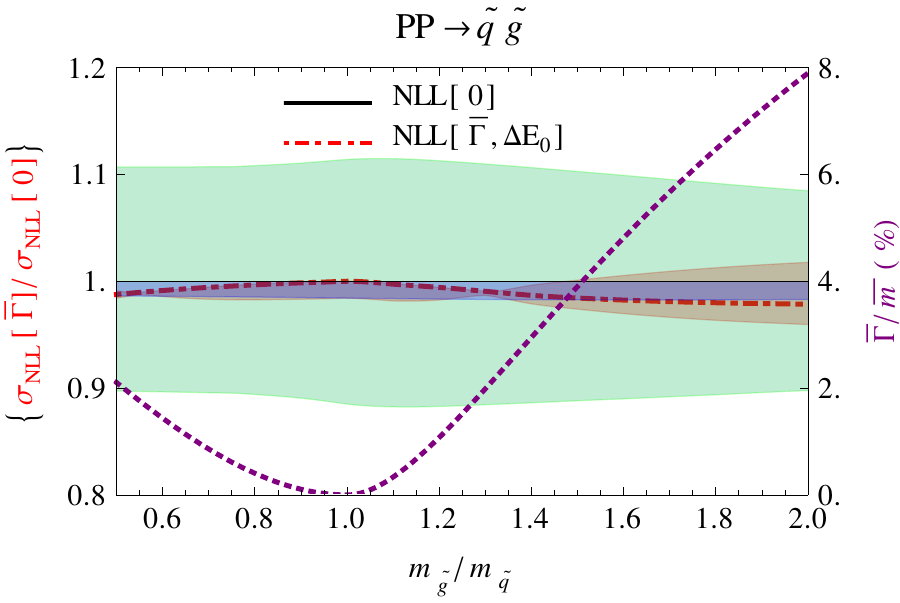}
\caption{Same as in Figure \ref{fig:NLLratio1}, but for the ratio $\sigma_{\text{NLL}}^{\text{matched},(2)}(\bar\Gamma)/\sigma_{\text{NLL}}^{\text{matched}}(0)$ of
  the matched cross section ~\eqref{eq:matchwidth2} including only
  finite-width effects beyond NLO  to 
the zero-width approximation~\eqref{eq:matchzero}. See Figure \ref{fig:NLLratio1} and the text for explanation.} \label{fig:NLLratio2}
\end{figure}
It can be seen that for all processes the finite-width result
including uncertainties lies within the uncertainty estimate of the
NLL calculation for stable squarks and gluinos if $\bar \Gamma/\bar
m\lesssim 5\%$.  For larger widths, large corrections are observed for
the gluino-gluino and squark-gluino processes, where the ratio
$\bar\Gamma/\bar m$ grows up to $12\%$, and the finite-width
corrections become of the order of $20\%$.  In this case the
uncertainties due to the treatment of finite-width effects are of a
similar magnitude as those due to scale and resummation ambiguities.
This indicates that corrections beyond the leading $E\rightarrow
E+i\bar\Gamma$ replacement, such as higher-order non-relativistic
corrections and non-resonant corrections become relevant.  A more
accurate treatment would require the matching of the NLL results to
the exact NLO QCD cross sections for the full $\tilde q\tilde q q$ or
$\tilde q\tilde q qq$-processes and the inclusion of the
N${}^{3/2}$LO corrections in the counting~\eqref{eq:syst} that are expected to be
of the order $\delta^{3/2}\sim (\bar\Gamma/\bar m)^{3/2}\sim 5\%$,
both of which are not available at the moment. Nevertheless it is
interesting to anticipate such a future matching to a full NLO
calculation by using the matching prescription~\eqref{eq:matchwidth2}
where the finite-width effects are included purely beyond NLO.
\begin{figure}[t!]
\centering
\includegraphics[width=0.49 \linewidth]{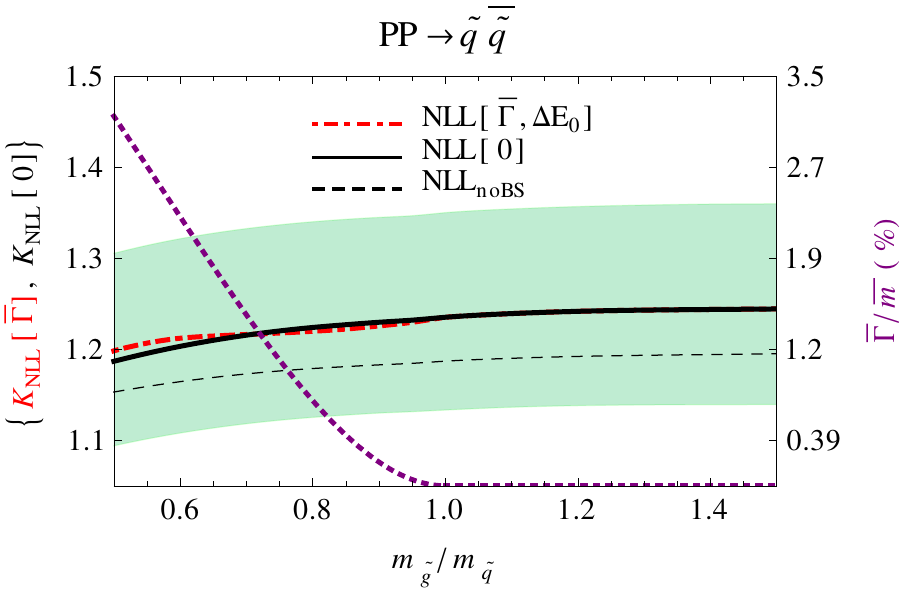}
\includegraphics[width=0.49 \linewidth]{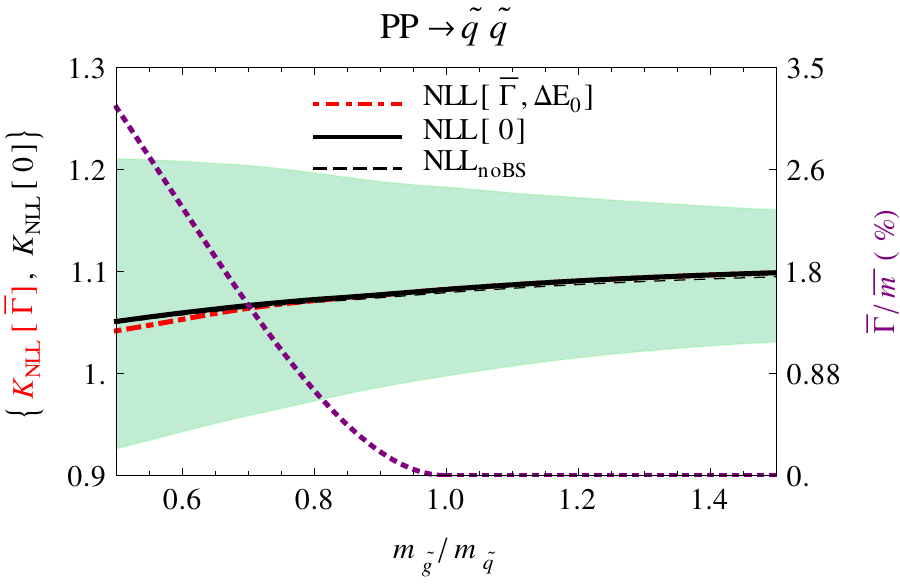}\\
\includegraphics[width=0.49 \linewidth]{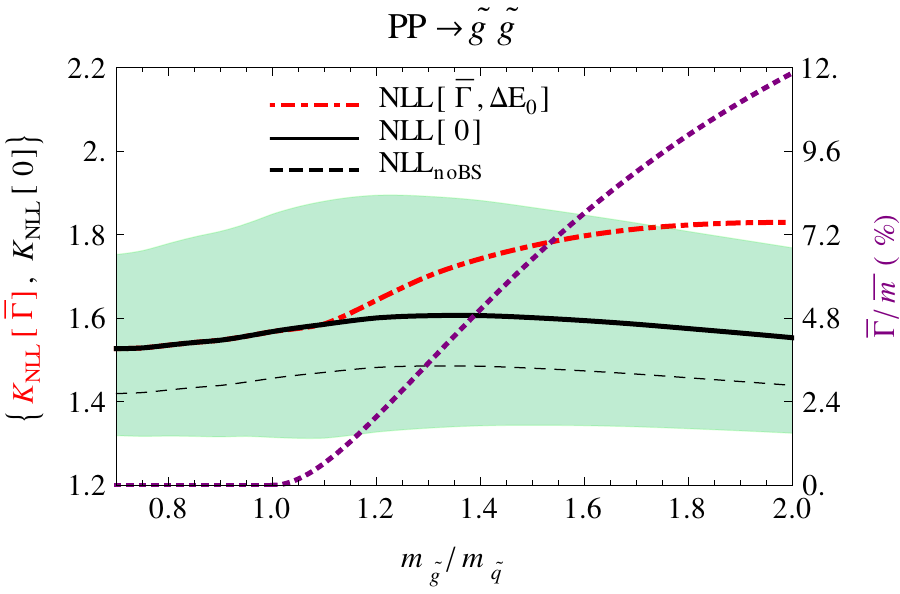}
\includegraphics[width=0.49 \linewidth]{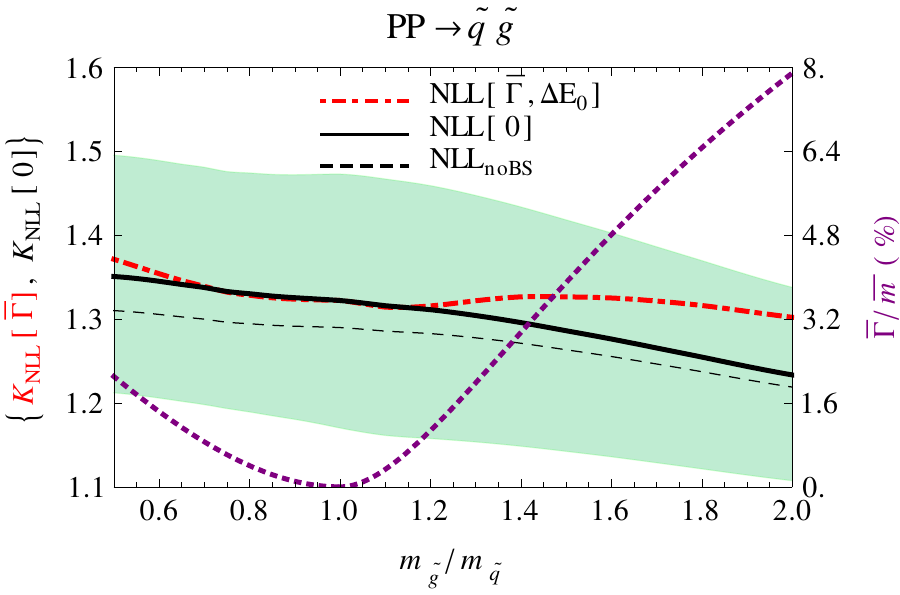}
\caption{The $K$-factors of the NLL cross sections including (red dot-dashed) and excluding (black) the widths for default $\Delta E=\Delta E_0$ and $\bar m=1500$~GeV at centre of mass energy 
$\sqrt{s}=8$ TeV. Dashed black: NLL resummed cross sections excluding bound-state contributions. Green band: Total error for zero width.} \label{fig:noBS}
\end{figure}
The results are shown in Figure \ref{fig:NLLratio2}, where the colour coding used for the lines is the same as in Figure \ref{fig:NLLratio1}. It is seen that the 
finite-width effects on the central values, as well as ambiguities due to the $\Delta E$ variation and the $E$- and $\beta$-implementation difference, are 
much smaller than for the matching~\eqref {eq:matchwidth}. In particular, the total uncertainty bands 
related to the finite-width implementation (red) lie now fully within the total zero-width NLL error bands (green). The difference of the plots in Figure \ref{fig:NLLratio1} 
and Figure \ref{fig:NLLratio2} indicates the potential impact 
of a full NLO SQCD calculation for the processes~\eqref{eq:decayprocesses} and shows that the finite-width effects beyond NLO are expected to be small.

To assess the size of the finite width corrections relative to the NLL soft and Coulomb corrections, in Figure \ref{fig:noBS} we compare the K-factors for our NLL result including finite decay widths to the result for stable squarks and gluinos~\eqref{eq:matchzero}. 
The $K$-factors relative to the NLO prediction are defined as 
$K_{\text{NLL}}=\sigma_{\text{NLL}}/\sigma_{\text{NLO}}$, where the finite-width approximation~\eqref{eq:NLOwidth} is always used for the denominator. 
We also show the result $\text{NLL}_{\text{noBS}}$ defined as the cross section in the stable case, excluding the bound-state contribution below threshold.
In general, for relatively narrow particles with $\bar\Gamma/\bar m\lesssim 5\%$ the K-factors including finite-width effects are well approximated by those for the zero-width case 
including the bound-state contribution, with the exception of
squark-gluino production and gluino-gluino production for the case $m_{\tilde g}>m_{\tilde q}$, where larger
finite-width effects are observed.

\begin{figure}[t!]
\centering
\includegraphics[width=0.49 \linewidth]{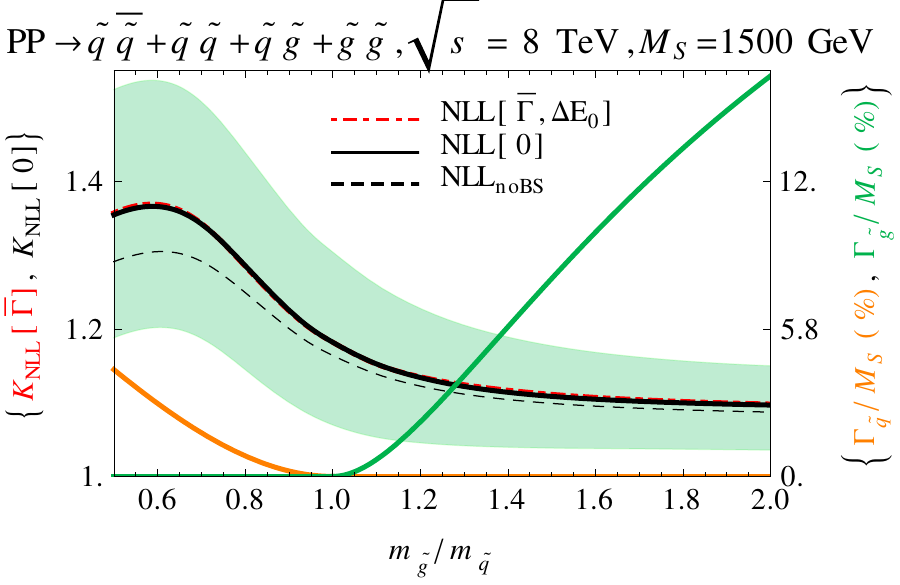}
\includegraphics[width=0.49 \linewidth]{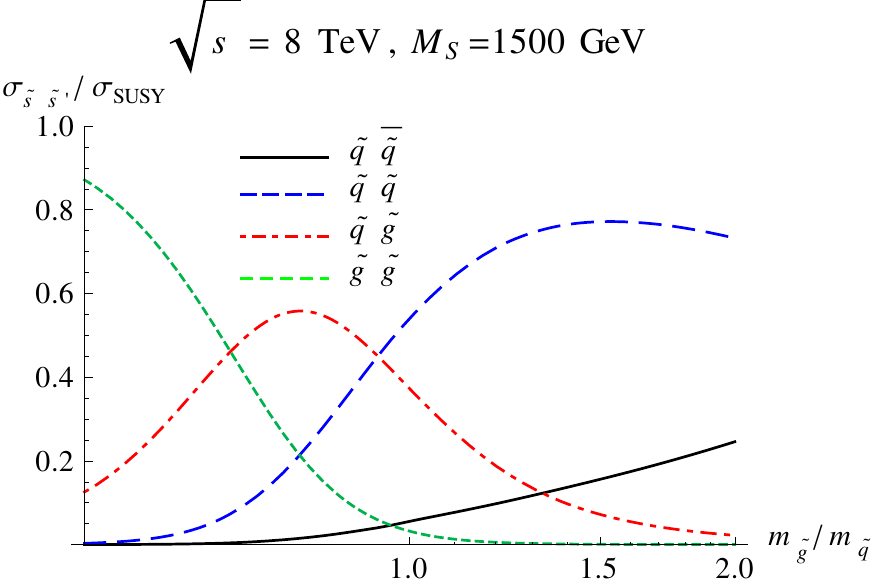}
\caption{Left plot: The NLL $K$-factor  including (dot-dashed red) and excluding (black) the widths for default $\Delta E=\Delta E_0$ and $M_S=(m_{\tilde{q}}+m_{\tilde{g}})/2=1500$~ GeV at 
centre of mass energy $\sqrt{s}=8$ TeV. Dashed black: NLL $K$-factor excluding bound-state contributions. Green band: Total error for zero width. The plot also shows the squark (thick orange) and gluino (thick green) widths (right 
vertical axis). Right plot: The ratio of the four relevant NLL cross sections to the total SUSY cross section.} \label{fig:SUSY}
\end{figure}
The relatively large finite-width corrections for gluino-production processes for $m_{\tilde g}>m_{\tilde q}$ have to be compared to the contributions of these processes to the total SUSY production rate.
 This can be seen in Figure \ref{fig:SUSY}, where in the 
left plot the $K_{\text{NLL}}$-factor is shown  for the finite-width and zero-width case. The effect of the QCD widths is seen to be negligible and the finite width can be safely set to zero when it comes to 
the total SUSY cross section. This is also expected since the numerically dominant processes are almost always those whose final-state particles are stable against SQCD decay, as can be seen from the right 
plot of Figure \ref{fig:SUSY}. 
 The inclusion of electroweak widths is not expected to change our qualitative results. These can be dominating when $r\simeq1$, but are only about the 
order of $\bar\Gamma/\bar m \sim 1\%$ and thus the effects are expected to be negligible.

\begin{figure}[t!]
\centering
\includegraphics[width=0.49 \linewidth]{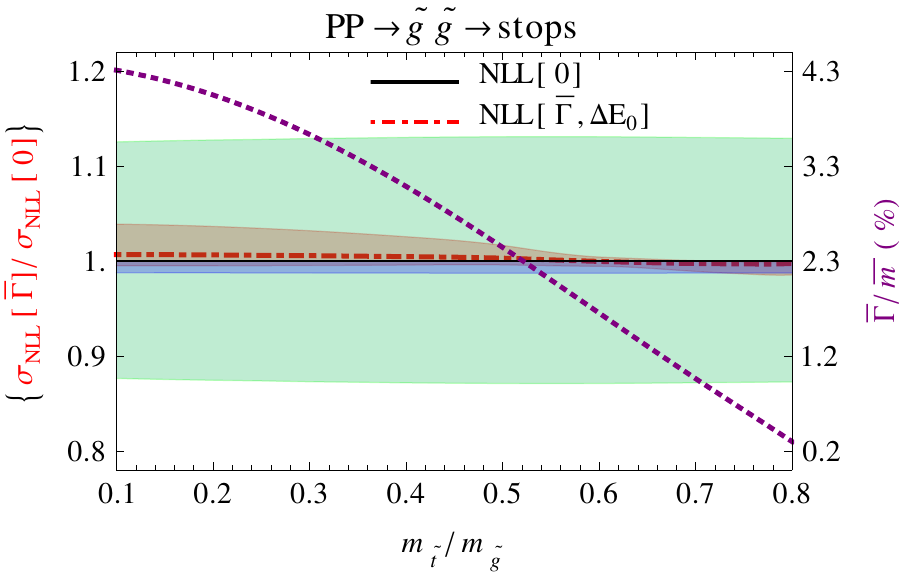}
\caption{ Ratio of the matched finite-width NLL cross
  section~\eqref{eq:matchwidth} to the zero-width
  approximation~\eqref{eq:matchzero} for gluino-pair production and
  subsequent decay to stops, as a function of
  $m_{\tilde{t}}/m_{\tilde{g}}$ for a fixed gluino mass
  $m_{\tilde{g}}=1000\,$GeV, a common light-flavour squark mass of
  $m_{\tilde{q}}=2000\,$GeV. The error bands for finite- and
  zero-width curves are defined as in Figures \ref{fig:NLLratio1} and
  \ref{fig:NLLratio2}.} \label{fig:ggtostop}
\end{figure}
As mentioned in Section~\ref{sec:frame}, in the study of finite-width effects considered in this work we have neglected the decay of gluinos to stops.
We do not expect the inclusion of this additional decay channel to significantly change the results presented in this section, except in the case in which 
gluino production is the dominant SUSY production channel and the gluinos only decay to stops. This is for example the case of scenarios with the
mass hierarchy $m_{\tilde{t}} < m_{\tilde{g}} \ll m_{\tilde{q}}$, with all squarks heavy except for stops. To have an estimate of the effects of a finite gluino
width in this particular situation in Figure \ref{fig:ggtostop} we show the NLL cross section for gluino-gluino production as a function of the ratio $m_{\tilde{t}}/m_{\tilde{g}}$
for a fixed gluino mass $m_{\tilde{g}}=1000\,$GeV, a common light-flavour squark mass of $m_{\tilde{q}}=2000\,$GeV and top mass $m_t=174\,$GeV. As in Figure \ref{fig:NLLratio1} 
the cross section is normalized to the gluino-pair production cross section at zero width, and the green band represents the total uncertainty of the
zero-width result, while the purple dashed curve gives the gluino width as a function of $m_{\tilde{t}}/m_{\tilde{g}}$. As can be seen from the plot the 
finite-width effects amount to a maximum of $\sim 1-2\%$ for small stop masses and the uncertainty associated with the finite-width result is contained
in the zero-width case error band. We thus expect corrections to the plots in Figures \ref{fig:noBS}, \ref{fig:SUSY} to amount to at most few percents.

\section{Conclusions}

In this work we have considered corrections related to finite widths of squarks and gluinos on the NLL resummed results presented in \cite{Falgari:2012hx}. 
As anticipated, a finite decay width leads to a screening of higher-order soft and Coulomb corrections. 
However we have found that for decay widths of the order of 
$\bar\Gamma/\bar m\lesssim 5\%$, corresponding to gluino to squark mass ratios of $0.5\lesssim m_{\tilde{g}}/m_{\tilde{q}}\lesssim 1.4$, finite-width effects on 
the NLL corrections are small and  within the uncertainty of the soft and Coulomb resummation including bound-state effects performed in \cite{Falgari:2012hx}. 

For larger decay widths of the order of $\bar\Gamma/\bar m\sim \alpha_s$ non-resonant contributions and higher order
non-relativistic corrections become important.  In this case it
becomes necessary to perform a calculation of order $\delta^{3/2}\sim
\alpha_s\sqrt{\bar\Gamma/\bar m}$ in unstable particle-effective
theory matched to an exact NLO calculation for three- or four-particle
final states including finite-width effects, which should be feasible
but challenging with current NLO technology. The accuracy of such a
calculation is expected to be better than $5\%$. For the
doubly-resonant contributions to the cross section, we have observed that finite-width corrections beyond NLO are small.

 Because the pair production processes of the lighter coloured sparticles, which are  stable with respect to SQCD, dominate the total SUSY cross section,
finite-width effects on the total SUSY production rate are negligible
on the whole range of squark and gluino widths considered.

\subsubsection*{Acknowledgements}
We are very grateful to C.~Speckner for his help with Whizard and to F.~Siegert for helping us out with SHERPA. 
The work of P.F. is supported by the ``Stichting voor Fundamenteel Onderzoek der Materie (FOM)", 
the work of C.W. by the research programme Mozaiek, which is partly financed by the Netherlands Organisation 
for Scientific Research (NWO).

\providecommand{\href}[2]{#2}\begingroup\raggedright\endgroup


\begin{thebibliography}{10}

\bibitem{Kulesza:2008jb}
A.~Kulesza and L.~Motyka,
  \href{http://dx.doi.org/10.1103/PhysRevLett.102.111802}{{\em Phys. Rev.
  Lett.} {\bf 102} (2009)  111802},
\href{http://arxiv.org/abs/0807.2405}{{\tt arXiv:0807.2405 [hep-ph]}}.

\bibitem{Kulesza:2009kq}
A.~Kulesza and L.~Motyka,
  \href{http://dx.doi.org/10.1103/PhysRevD.80.095004}{{\em Phys. Rev.} {\bf
  D80} (2009)  095004},
\href{http://arxiv.org/abs/0905.4749}{{\tt arXiv:0905.4749 [hep-ph]}}.

\bibitem{Beenakker:2009ha}
W.~Beenakker {\em et al.},
  \href{http://dx.doi.org/10.1088/1126-6708/2009/12/041}{{\em JHEP} {\bf 12}
  (2009)  041},
\href{http://arxiv.org/abs/0909.4418}{{\tt arXiv:0909.4418 [hep-ph]}}.

\bibitem{Beenakker:2010nq}
W.~Beenakker, S.~Brensing, M.~Kr{\"a}mer, A.~Kulesza, E.~Laenen, {\em et al.},
  \href{http://dx.doi.org/10.1007/JHEP08(2010)098}{{\em JHEP} {\bf 1008} (2010)
   098}, \href{http://arxiv.org/abs/1006.4771}{{\tt arXiv:1006.4771 [hep-ph]}}.

\bibitem{Beenakker:2011sf}
W.~Beenakker, S.~Brensing, M.~Kr{\"a}mer, A.~Kulesza, E.~Laenen, {\em et al.},
  {\em JHEP} {\bf 1201} (2012)  076,
\href{http://arxiv.org/abs/1110.2446}{{\tt arXiv:1110.2446 [hep-ph]}}.

\bibitem{Langenfeld:2009eg}
U.~Langenfeld and S.-O. Moch, {\em Phys. Lett.} {\bf B675} (2009)  210--221,
\href{http://arxiv.org/abs/0901.0802}{{\tt arXiv:0901.0802 [hep-ph]}}.

\bibitem{Langenfeld:2010vu}
U.~Langenfeld, \href{http://dx.doi.org/10.1007/JHEP07(2011)052}{{\em JHEP} {\bf
  1107} (2011)  052},
\href{http://arxiv.org/abs/1011.3341}{{\tt arXiv:1011.3341 [hep-ph]}}.

\bibitem{Langenfeld:2012ti}
U.~Langenfeld, S.-O. Moch, and T.~Pfoh,
\href{http://arxiv.org/abs/1208.4281}{{\tt arXiv:1208.4281 [hep-ph]}}.

\bibitem{Beneke:2009rj}
M.~Beneke, P.~Falgari, and C.~Schwinn,
  \href{http://dx.doi.org/10.1016/j.nuclphysb.2009.11.004}{{\em Nucl. Phys.}
  {\bf B828} (2010)  69--101},
\href{http://arxiv.org/abs/0907.1443}{{\tt arXiv:0907.1443 [hep-ph]}}.

\bibitem{Beneke:2009nr}
M.~Beneke, P.~Falgari, and C.~Schwinn, {\em PoS} {\bf (EPS-HEP 2009)} (2009)
  319,
\href{http://arxiv.org/abs/0909.3488}{{\tt arXiv:0909.3488 [hep-ph]}}.

\bibitem{Beneke:2010da}
M.~Beneke, P.~Falgari, and C.~Schwinn,
  \href{http://dx.doi.org/10.1016/j.nuclphysb.2010.09.009}{{\em Nucl. Phys.}
  {\bf B842} (2011)  },
\href{http://arxiv.org/abs/1007.5414}{{\tt arXiv:1007.5414 [hep-ph]}}.

\bibitem{Falgari:2012hx}
P.~Falgari, C.~Schwinn, and C.~Wever,
  \href{http://dx.doi.org/10.1007/JHEP06(2012)052}{{\em JHEP} {\bf 1206} (2012)
   052},
\href{http://arxiv.org/abs/1202.2260}{{\tt arXiv:1202.2260 [hep-ph]}}.

\bibitem{Sterman:1986aj}
G.~Sterman,
{\em Nucl. Phys.} {\bf B281} (1987)  310.

\bibitem{Catani:1989ne}
S.~Catani and L.~Trentadue,
{\em Nucl. Phys.} {\bf B327} (1989)  323.

\bibitem{Kidonakis:1997gm}
N.~Kidonakis and G.~Sterman,
  \href{http://dx.doi.org/10.1016/S0550-3213(97)00506-3}{{\em Nucl. Phys.} {\bf
  B505} (1997)  321--348},
\href{http://arxiv.org/abs/hep-ph/9705234}{{\tt arXiv:hep-ph/9705234}}.

\bibitem{Bonciani:1998vc}
R.~Bonciani, S.~Catani, M.~L. Mangano, and P.~Nason,
  \href{http://dx.doi.org/10.1016/S0550-3213(98)00335-6}{{\em Nucl. Phys.} {\bf
  B529} (1998)  424--450},
\href{http://arxiv.org/abs/hep-ph/9801375}{{\tt arXiv:hep-ph/9801375}}.

\bibitem{Fadin:1987wz}
V.~S. Fadin and V.~A. Khoze,
{\em JETP Lett.} {\bf 46} (1987)  525--529.

\bibitem{Kauth:2009ud}
M.~R. Kauth, J.~H. K{\"u}hn, P.~Marquard, and M.~Steinhauser,
  \href{http://dx.doi.org/10.1016/j.nuclphysb.2010.01.019}{{\em Nucl.Phys.}
  {\bf B831} (2010)  285--305},
\href{http://arxiv.org/abs/0910.2612}{{\tt arXiv:0910.2612 [hep-ph]}}.

\bibitem{Younkin:2009zn}
J.~E. Younkin and S.~P. Martin,
  \href{http://dx.doi.org/10.1103/PhysRevD.81.055006}{{\em Phys.Rev.} {\bf D81}
  (2010)  055006}, \href{http://arxiv.org/abs/0912.4813}{{\tt arXiv:0912.4813
  [hep-ph]}}.

\bibitem{Bigi:1991mi}
I.~I.~Y. Bigi, V.~S. Fadin, and V.~A. Khoze,
\href{http://dx.doi.org/10.1016/0550-3213(92)90297-O}{{\em Nucl. Phys.} {\bf
  B377} (1992)  461--479}.

\bibitem{Hagiwara:2009hq}
K.~Hagiwara and H.~Yokoya,
  \href{http://dx.doi.org/10.1088/1126-6708/2009/10/049}{{\em JHEP} {\bf 10}
  (2009)  049},
\href{http://arxiv.org/abs/0909.3204}{{\tt arXiv:0909.3204 [hep-ph]}}.

\bibitem{Kauth:2011vg}
M.~R. Kauth, J.~H. K{\"u}hn, P.~Marquard, and M.~Steinhauser,
  \href{http://dx.doi.org/10.1016/j.nuclphysb.2011.11.024}{{\em Nucl.Phys.}
  {\bf B857} (2012)  28--64},
\href{http://arxiv.org/abs/1108.0361}{{\tt arXiv:1108.0361 [hep-ph]}}.

\bibitem{Kauth:2011bz}
M.~R. Kauth, A.~Kress, and J.~H. K{\"u}hn,
  \href{http://dx.doi.org/10.1007/JHEP12(2011)104}{{\em JHEP} {\bf 1112} (2011)
   104},
\href{http://arxiv.org/abs/1108.0542}{{\tt arXiv:1108.0542 [hep-ph]}}.

\bibitem{Hagiwara:2005wg}
K.~Hagiwara, W.~Kilian, F.~Krauss, T.~Ohl, T.~Plehn, {\em et al.},
  \href{http://dx.doi.org/10.1103/PhysRevD.73.055005}{{\em Phys.Rev.} {\bf D73}
  (2006)  055005},
\href{http://arxiv.org/abs/hep-ph/0512260}{{\tt arXiv:hep-ph/0512260
  [hep-ph]}}.

\bibitem{Berdine:2007uv}
D.~Berdine, N.~Kauer, and D.~Rainwater,
  \href{http://dx.doi.org/10.1103/PhysRevLett.99.111601}{{\em Phys.Rev.Lett.}
  {\bf 99} (2007)  111601},
\href{http://arxiv.org/abs/hep-ph/0703058}{{\tt arXiv:hep-ph/0703058
  [hep-ph]}}.

\bibitem{Uhlemann:2008pm}
C.~Uhlemann and N.~Kauer,
  \href{http://dx.doi.org/10.1016/j.nuclphysb.2009.01.022}{{\em Nucl.Phys.}
  {\bf B814} (2009)  195--211},
\href{http://arxiv.org/abs/0807.4112}{{\tt arXiv:0807.4112 [hep-ph]}}.

\bibitem{Gigg:2008yc}
M.~Gigg and P.~Richardson,
\href{http://arxiv.org/abs/0805.3037}{{\tt arXiv:0805.3037 [hep-ph]}}.

\bibitem{Hollik:2012rc}
W.~Hollik, J.~M. Lindert, and D.~Pagani,
\href{http://arxiv.org/abs/1207.1071}{{\tt arXiv:1207.1071 [hep-ph]}}.

\bibitem{Beneke:2003xh}
M.~Beneke, A.~P. Chapovsky, A.~Signer, and G.~Zanderighi,
  \href{http://dx.doi.org/10.1103/PhysRevLett.93.011602}{{\em Phys. Rev. Lett.}
  {\bf 93} (2004)  011602},
\href{http://arxiv.org/abs/hep-ph/0312331}{{\tt arXiv:hep-ph/0312331}}.

\bibitem{Beneke:2004km}
M.~Beneke, A.~P. Chapovsky, A.~Signer, and G.~Zanderighi,
  \href{http://dx.doi.org/10.1016/j.nuclphysb.2004.03.016}{{\em Nucl. Phys.}
  {\bf B686} (2004)  205--247},
\href{http://arxiv.org/abs/hep-ph/0401002}{{\tt arXiv:hep-ph/0401002}}.

\bibitem{Beneke:2007zg}
M.~Beneke, P.~Falgari, C.~Schwinn, A.~Signer, and G.~Zanderighi,
  \href{http://dx.doi.org/10.1016/j.nuclphysb.2007.09.030}{{\em Nucl. Phys.}
  {\bf B792} (2008)  89--135},
\href{http://arxiv.org/abs/0707.0773}{{\tt arXiv:0707.0773 [hep-ph]}}.

\bibitem{Actis:2008rb}
S.~Actis, M.~Beneke, P.~Falgari, and C.~Schwinn,
  \href{http://dx.doi.org/10.1016/j.nuclphysb.2008.08.006}{{\em Nucl. Phys.}
  {\bf B807} (2009)  1--32},
\href{http://arxiv.org/abs/0807.0102}{{\tt arXiv:0807.0102 [hep-ph]}}.

\bibitem{Beenakker:1996dw}
W.~Beenakker, R.~H{\"o}pker, and P.~Zerwas,
  \href{http://dx.doi.org/10.1016/0370-2693(96)00379-6}{{\em Phys.Lett.} {\bf
  B378} (1996)  159--166},
\href{http://arxiv.org/abs/hep-ph/9602378}{{\tt arXiv:hep-ph/9602378
  [hep-ph]}}.

\bibitem{Baer:1986au}
H.~Baer, V.~D. Barger, D.~Karatas, and X.~Tata,
\href{http://dx.doi.org/10.1103/PhysRevD.36.96}{{\em Phys.Rev.} {\bf D36}
  (1987)  96}.

\bibitem{Baer:1990sc}
H.~Baer, X.~Tata, and J.~Woodside,
\href{http://dx.doi.org/10.1103/PhysRevD.42.1568}{{\em Phys.Rev.} {\bf D42}
  (1990)  1568--1576}.

\bibitem{Porod:2003um}
W.~Porod, \href{http://dx.doi.org/10.1016/S0010-4655(03)00222-4}{{\em
  Comput.Phys.Commun.} {\bf 153} (2003)  275--315},
\href{http://arxiv.org/abs/hep-ph/0301101}{{\tt arXiv:hep-ph/0301101
  [hep-ph]}}.

\bibitem{Porod:2011nf}
W.~Porod and F.~Staub, \href{http://dx.doi.org/10.1016/j.cpc.2012.05.021}{{\em
  Comput.Phys.Commun.} {\bf 183} (2012)  2458--2469},
\href{http://arxiv.org/abs/1104.1573}{{\tt arXiv:1104.1573 [hep-ph]}}.

\bibitem{Muhlleitner:2003vg}
M.~M{\"u}hlleitner, A.~Djouadi, and Y.~Mambrini,
  \href{http://dx.doi.org/10.1016/j.cpc.2005.01.012}{{\em Comput.Phys.Commun.}
  {\bf 168} (2005)  46--70},
\href{http://arxiv.org/abs/hep-ph/0311167}{{\tt arXiv:hep-ph/0311167
  [hep-ph]}}.

\bibitem{Djouadi:2006bz}
A.~Djouadi, M.~M{\"u}hlleitner, and M.~Spira, {\em Acta Phys.Polon.} {\bf B38}
  (2007)  635--644,
\href{http://arxiv.org/abs/hep-ph/0609292}{{\tt arXiv:hep-ph/0609292
  [hep-ph]}}.

\bibitem{Denner:2010jp}
A.~Denner, S.~Dittmaier, S.~Kallweit, and S.~Pozzorini,
  \href{http://dx.doi.org/10.1103/PhysRevLett.106.052001}{{\em Phys. Rev.
  Lett.} {\bf 106} (2011)  052001},
\href{http://arxiv.org/abs/1012.3975}{{\tt arXiv:1012.3975 [hep-ph]}}.

\bibitem{Bevilacqua:2010qb}
G.~Bevilacqua, M.~Czakon, A.~van Hameren, C.~G. Papadopoulos, and M.~Worek,
  \href{http://dx.doi.org/10.1007/JHEP02(2011)083}{{\em JHEP} {\bf 02} (2011)
  083},
\href{http://arxiv.org/abs/1012.4230}{{\tt arXiv:1012.4230 [hep-ph]}}.

\bibitem{GoncalvesNetto:2012yt}
D.~Goncalves-Netto, D.~Lopez-Val, K.~Mawatari, T.~Plehn, and I.~Wigmore,
\href{http://arxiv.org/abs/1211.0286}{{\tt arXiv:1211.0286 [hep-ph]}}.

\bibitem{Stuart:1991xk}
R.~G. Stuart,
{\em Phys. Lett.} {\bf B262} (1991)  113--119.

\bibitem{Aeppli:1993rs}
A.~Aeppli, G.~J. van Oldenborgh, and D.~Wyler, {\em Nucl. Phys.} {\bf B428}
  (1994)  126--146,
\href{http://arxiv.org/abs/hep-ph/9312212}{{\tt hep-ph/9312212}}.

\bibitem{Fadin:1993dz}
V.~S. Fadin, V.~A. Khoze, and A.~D. Martin,
{\em Phys. Rev.} {\bf D49} (1994)  2247--2256.

\bibitem{Melnikov:1993np}
K.~Melnikov and O.~I. Yakovlev,
  \href{http://dx.doi.org/10.1016/0370-2693(94)90410-3}{{\em Phys.Lett.} {\bf
  B324} (1994)  217--223},
\href{http://arxiv.org/abs/hep-ph/9302311}{{\tt arXiv:hep-ph/9302311
  [hep-ph]}}.

\bibitem{Beenakker:1996ch}
W.~Beenakker, R.~H{\"o}pker, M.~Spira, and P.~M. Zerwas,
  \href{http://dx.doi.org/10.1016/S0550-3213(97)00084-9}{{\em Nucl. Phys.} {\bf
  B492} (1997)  51--103},
\href{http://arxiv.org/abs/hep-ph/9610490}{{\tt arXiv:hep-ph/9610490}}.

\bibitem{Bauer:2000yr}
C.~W. Bauer, S.~Fleming, D.~Pirjol, and I.~W. Stewart, {\em Phys. Rev.} {\bf
  D63} (2001)  114020,
\href{http://arxiv.org/abs/hep-ph/0011336}{{\tt hep-ph/0011336}}.

\bibitem{Bauer:2001yt}
C.~W. Bauer, D.~Pirjol, and I.~W. Stewart, {\em Phys. Rev.} {\bf D65} (2002)
  054022,
\href{http://arxiv.org/abs/hep-ph/0109045}{{\tt hep-ph/0109045}}.

\bibitem{Beneke:2002ph}
M.~Beneke, A.~P. Chapovsky, M.~Diehl, and T.~Feldmann, {\em Nucl. Phys.} {\bf
  B643} (2002)  431--476,
\href{http://arxiv.org/abs/hep-ph/0206152}{{\tt hep-ph/0206152}}.

\bibitem{Brambilla:1999xf}
N.~Brambilla, A.~Pineda, J.~Soto, and A.~Vairo,
  \href{http://dx.doi.org/10.1016/S0550-3213(99)00693-8}{{\em Nucl. Phys.} {\bf
  B566} (2000)  275},
\href{http://arxiv.org/abs/hep-ph/9907240}{{\tt arXiv:hep-ph/9907240}}.

\bibitem{Beneke:2010mp}
M.~Beneke, B.~Jantzen, and P.~Ruiz-Femenia,
  \href{http://dx.doi.org/10.1016/j.nuclphysb.2010.07.006}{{\em Nucl.Phys.}
  {\bf B840} (2010)  186--213}, \href{http://arxiv.org/abs/1004.2188}{{\tt
  arXiv:1004.2188 [hep-ph]}}.

\bibitem{Penin:2011gg}
A.~A. Penin and J.~H. Piclum,
  \href{http://dx.doi.org/10.1007/JHEP01(2012)034}{{\em JHEP} {\bf 1201} (2012)
   034},
\href{http://arxiv.org/abs/1110.1970}{{\tt arXiv:1110.1970 [hep-ph]}}.

\bibitem{Becher:2009kw}
T.~Becher and M.~Neubert,
  \href{http://dx.doi.org/10.1103/PhysRevD.79.125004}{{\em Phys. Rev.} {\bf
  D79} (2009)  125004},
\href{http://arxiv.org/abs/0904.1021}{{\tt arXiv:0904.1021 [hep-ph]}}.

\bibitem{Czakon:2009zw}
M.~Czakon, A.~Mitov, and G.~Sterman,
  \href{http://dx.doi.org/10.1103/PhysRevD.80.074017}{{\em Phys. Rev.} {\bf
  D80} (2009)  074017},
\href{http://arxiv.org/abs/0907.1790}{{\tt arXiv:0907.1790 [hep-ph]}}.

\bibitem{Becher:2006nr}
T.~Becher and M.~Neubert, {\em Phys. Rev. Lett.} {\bf 97} (2006)  082001,
\href{http://arxiv.org/abs/hep-ph/0605050}{{\tt hep-ph/0605050}}.

\bibitem{Wichmann:1961}
E.~H. Wichmann and C.-H. Woo, \href{http://dx.doi.org/10.1063/1.1703696}{{\em
  Journal of Mathematical Physics} {\bf 2} (1961) no.~2, 178--180}.
  \url{http://link.aip.org/link/?JMP/2/178/1}.

\bibitem{Beneke:1999zr}
M.~Beneke, \href{http://arxiv.org/abs/hep-ph/9911490}{{\tt
  arXiv:hep-ph/9911490}}.
Proceedings of the 8th International Symposium on Heavy Flavor Physics (Heavy
  Flavors 8), Southampton, England, 25-29 Jul 1999.

\bibitem{Kats:2009bv}
Y.~Kats and M.~D. Schwartz,
  \href{http://dx.doi.org/10.1007/JHEP04(2010)016}{{\em JHEP} {\bf 04} (2010)
  016},
\href{http://arxiv.org/abs/0912.0526}{{\tt arXiv:0912.0526 [hep-ph]}}.

\bibitem{Hoang:2010gu}
A.~H. Hoang, C.~J. Reisser, and P.~Ruiz-Femenia,
  \href{http://dx.doi.org/10.1103/PhysRevD.82.014005}{{\em Phys.Rev.} {\bf D82}
  (2010)  014005}, \href{http://arxiv.org/abs/1002.3223}{{\tt arXiv:1002.3223
  [hep-ph]}}.

\bibitem{RuizFemenia:2012ma}
P.~Ruiz-Femenia, \href{http://arxiv.org/abs/1203.0934}{{\tt arXiv:1203.0934
  [hep-ph]}}.
Contribution to the proceedings of the 2011 International Workshop on Future
  Linear Colliders (LCWS11), Sept. 26-30, Granada, Spain.

\bibitem{Kilian:2007gr}
W.~Kilian, T.~Ohl, and J.~Reuter,
  \href{http://dx.doi.org/10.1140/epjc/s10052-011-1742-y}{{\em Eur.Phys.J.}
  {\bf C71} (2011)  1742},
\href{http://arxiv.org/abs/0708.4233}{{\tt arXiv:0708.4233 [hep-ph]}}.

\bibitem{Gleisberg:2008ta}
T.~Gleisberg, S.~H{\"o}che, F.~Krauss, M.~Sch{\"o}nherr, S.~Schumann, {\em et
  al.}, \href{http://dx.doi.org/10.1088/1126-6708/2009/02/007}{{\em JHEP} {\bf
  0902} (2009)  007},
\href{http://arxiv.org/abs/0811.4622}{{\tt arXiv:0811.4622 [hep-ph]}}.

\bibitem{Beenakker:1996ed}
W.~Beenakker, R.~H{\"o}pker, and M.~Spira,
\href{http://arxiv.org/abs/hep-ph/9611232}{{\tt arXiv:hep-ph/9611232}}.

\end{thebibliography}
\end{document}